\documentclass[prx,amsmath,amssymb,amsfonts,superscriptaddress,twocolumn,aps,10pt]{revtex4-2}
\usepackage{graphicx,color,mathtools,bm,braket,newtxtext,newtxmath,physics,csquotes,siunitx,comment}
\usepackage[english]{babel}
\usepackage[T1]{fontenc}
\usepackage[utf8]{inputenc}
\usepackage[hyphenbreaks]{breakurl}
\usepackage[colorlinks=true,linkcolor=blue,citecolor=blue,urlcolor=blue]{hyperref}
\usepackage[normalem]{ulem}
\usepackage[acronym]{glossaries}

\DeclareUnicodeCharacter{03BC}{$\mu$}
\DeclareUnicodeCharacter{03BB}{$\lambda$}

\makeatletter
\newcommand*{\glsplainhyperlink}[2]{%
    \begingroup%
      \hypersetup{hidelinks}%
      \hyperlink{#1}{#2}%
    \endgroup%
}
\let\@glslink\glsplainhyperlink
\makeatother

\newacronym{kkd}{KKD}{Kramers-Kronig detection}
\newacronym{mp}{MP}{minimal phase}
\newacronym{ssb}{SSB}{single sideband}
\newacronym{hd}{HD}{homodyne detection}
\newacronym{dhd}{DHD}{double homodyne detection}
\newacronym{hrd}{HRD}{heterodyne detection}
\newacronym{lo}{LO}{local oscillator}
\newacronym{cvqkd}{CVQKD}{continuous variable quantum key distribution}
\newacronym{snr}{SNR}{signal-to-noise-ratio}
\newacronym{eom}{EOM}{electro-optic modulator}

\newcommand{\stirling}[2]{\genfrac{\{}{\}}{0pt}{}{#1}{#2}}
\newcommand{\func}[1]{\left[#1\right]}
\newcommand{\dif}{\mathrm{d}}
\newcommand{\vacuum}{0}
\newcommand{\inner}[2]{\left<#1,#2\right>}
\newcommand{\hd}{\mathrm{\gls[hyper=false]{hd}}}
\newcommand{\dhd}{\mathrm{\gls[hyper=false]{dhd}}}
\newcommand{\snr}{\mathrm{\gls[hyper=false]{snr}}}
\DeclareMathOperator{\fourier}{\mathcal{F}}
\DeclareMathOperator{\pval}{\mathcal{P}}
\newcommand{\kk}{\mathrm{\gls[hyper=false]{kkd}}}
\newcommand{\cohtket}[1]{\ket{\func{#1}}}
\newcommand{\hc}{\mathrm{h.c.}}
\DeclareMathOperator{\argument}{\mathrm{arg}}
\newcommand{\Dif}{\mathcal{D}}
\DeclareMathOperator{\pfunc}{\mathrm{P}}
\newcommand{\cohtbra}[1]{\bra{\func{#1}}}
\newcommand{\phket}[1]{\ket{#1}}
\newcommand{\eom}{\mathrm{\gls[hyper=false]{eom}}}
\DeclareMathOperator{\bessel}{\mathrm{J}}
\newcommand{\hilbert}{\mathcal{H}}
\newcommand{\fock}{\mathcal{F}}
\newcommand{\cc}{\mathrm{c.c.}}
\newcommand{\latin}[1]{\emph{#1}}
\newcommand{\filter}{\mathrm{filter}}

\begin{document}

\author{Thomas Pousset}
\email{thomas.pousset@telecom-paris.fr}
\affiliation{Telecom Paris, Institut Polytechnique de Paris, 19 Place Marguerite Perey, 91120 Palaiseau, France}
\author{Maxime Federico}
\email{maxime.federico@telecom-paris.fr}
\affiliation{Telecom Paris, Institut Polytechnique de Paris, 19 Place Marguerite Perey, 91120 Palaiseau, France}
\author{Romain Alléaume}
\email{romain.alleaume@telecom-paris.fr}
\affiliation{Telecom Paris, Institut Polytechnique de Paris, 19 Place Marguerite Perey, 91120 Palaiseau, France}
\author{Nicolas Fabre}
\email{nicolas.fabre@telecom-paris.fr}
\affiliation{Telecom Paris, Institut Polytechnique de Paris, 19 Place Marguerite Perey, 91120 Palaiseau, France}

\date{\today}

\begin{abstract}
We investigate the quantization of Kramers-Kronig detection technique initially developped for classical optical communications.
It consists in mixing the unknown field with a strong monochromatic local oscillator on an unbalanced beamsplitter.
A single output of the beamsplitter undergoes a direct detection of the optical intensity by means of a single photodiode.
When the measured output verifies signal processing constraints, namely, the minimal phase and the single sideband constraints, Kramers-Kronig detection reconstructs the phase of the signal from the intensity measurements via a digitally computed Hilbert transform.
The local oscillator being known, Kramers-Kronig detection allows for reconstructing the quadratures of the unknown field.
We show that this result holds in the quantum regime up to first order in the local oscillator amplitude and thus that Kramers-Kronig detection acts as a coherent detection able to measure both quadratures, making it a Gaussian measurement similar to double homodyne detection.
We also study in details the phase information measured by Kramers-Kronig detection for bosonic coherent states, monomode pure states and mixed states.
Finally, we propose and investigate a spectral tomography protocol for single-photon states that is inspired by Kramers-Kronig detection and relies on a spectral engineering of the single-photon.
\end{abstract}

\title{Kramers-Kronig detection in the quantum regime} 

\maketitle

\section{Introduction}\label{sec:intro}
Classical communication links can be decomposed into five parts according to the Shannon model~\cite{shannon_1948}: the encoder, the transmitter, the channel, the receiver, and the decoder.
The logical information is encoded into a physical message transmitted through a channel.
The receiver compensates for the channel effects to retrieve the transmitted message, which is then decoded.
In optical communications, the message is encoded into an electric field that propagates through an underwater optical fiber for subsea communications or through air for free space communications for example and is detected by either a coherent or a direct detector.
Direct detection measures only the intensity of the optical field, whereas coherent detection measures both the intensity and phase of the optical field.
Thus, for the same bandwidth, direct detection~\cite{ yan_2013,zhong_2018} captures less information than coherent detection.
However, coherent detectors are more expensive and less compact than direct detectors because they require more optical components, such as a local oscillator and balanced photodiodes, while traditional direct detectors such as intensity modulation-direct detection~\cite{yan_2013} systems only need a single photodiode.
Therefore, direct detection schemes are mostly used for short communication links in environments unsuited for stable operations, while coherent detection schemes are used for long-distance and high-rate communications~\cite{kikuchi_fundamentals_2016,zhong_2018}.
\par The most used coherent detections are balanced detections such as the \gls[hyper=false]{hd}~\cite{malyon_digital_1984}, the \gls[hyper=false]{dhd} also known as the phase diversity detection~\cite{kikuchi_phase-diversity_2006,ly-gagnon_coherent_2006} and the \gls[hyper=false]{hrd}~\cite{delange_1968}. 
The first reconstructs a single quadrature of the field by coupling the signal to be measured with a local oscillator at the same carrier frequency. 
The second reconstructs both quadratures by performing two \gls[hyper=false]{hd} at the cost of an attenuation of the signal.
The last one reconstructs both quadratures by downconverting the signal from the optical domain to the radiofrequency domain by coupling it with a local oscillator shifted in frequency at the cost of a \gls[hyper=false]{snr} reduction.
Recently a promising alternative to traditional coherent detectors and direct detectors has been proposed under the name of \gls[hyper=false]{kkd}~\cite{mecozzi_kramerskronig_2016,mecozzi_kramerskronig_2019}. 
It is a direct detection that acts as a coherent detection since both quadratures of the field can be reconstructed.
In the classical regime, \gls[hyper=false]{kkd} may be used for high speed wireless communications~\cite{harter_generalized_2020}, for intra-datacenter compact wireless-fiber communications~\cite{tavakkolnia_terabit_2022}, for datacenters interconnections~\cite{chen_single-wavelength_2017,chen_kramerskronig_2018} as well as outdoor free-space communications~\cite{lorences-riesgo_200_2020}.
\gls[hyper=false]{kkd} consists first in mixing a \gls[hyper=false]{lo} with a spectrally engineered signal on a beamsplitter.
A single output is measured and when it verifies the minimum phase condition~\cite{mecozzi_necessary_2016}, \gls[hyper=false]{kkd} reconstructs its relative phase compared to the \gls[hyper=false]{lo} by digitally processing the measured field intensity and recovers both quadratures.
\gls[hyper=false]{kkd} highly relies on digital signal processing instead of optical engineering and shifts the optical complexity into a signal processing complexity to retrieve the phase of the signal~\cite{zhong_2018,orsuti_edge-carrier-assisted_2023,ma_2024}.
\par In practice, when measuring the quadratures, several noises have to be taken into account such as the quantum fluctuations, the thermal noise and the phase noise for example.
In the quantum regime, namely when the \gls[hyper=false]{snr} is low, the main noise is due to quantum fluctuations   and thus the electromagnetic field and the detection must be described by using the quantum optics formalism.
Coherent detections~\cite{kikuchi_fundamentals_2016,banaszek_quantum_2020} such as \gls[hyper=false]{hd}~\cite{shapiro_optical_1979, yuen_optical_1980,yuen_noise_1983,banaszek_operational_1997}, \gls[hyper=false]{dhd}~\cite{walker_multiport_1986} and \gls[hyper=false]{hrd}~\cite{personick_1971a,yuen_noise_1983} have been thoroughly analysed in this regime. 
They are used in many quantum technology protocols such as quantum random number generators~\cite{gehring_homodyne-based_2021,bruynsteen_100_2023}, quantum tomography~\cite{smithey_measurement_1993,breitenbach_squeezed_1995,ourjoumtsev_2006}, quantum sensing~\cite{delaubert_tem10_2006,pinel_2012,jianRealtimeDisplacementMeasurement2012} or continuous variable quantum key distribution~\cite{madsen_2012,jouguet_2013,samsonov_2021,aymeric_symbiotic_2022,jain_2022,schiavon_2023, qi_2023,roumestan_2024}.
\par For \gls[hyper=false]{kkd}, quantum fluctuations have been calculated in~\cite{zhang_quantum_2023} assuming that it measures both quadratures of the field.
The quantum fluctuations associated with the measurement of a symbol using \gls[hyper=false]{kkd} are reported to be smaller compared to those of \gls[hyper=false]{hrd}, and by extension, \gls[hyper=false]{dhd}, as these two have the same SNR~\cite{kikuchi_fundamentals_2016}.
In Ref.~\cite{zhang_quantum_2023}, the symbols are carried by plane waves instead of normalized spectral modes.
In addition, only a finite number of plane waves is taken into account in the decomposition of the bosonic operators.
The consequence is that the symbol can be reconstructed by multiplying the reconstructed signal with the complex conjugated plane wave instead of projecting on a mode with an integral as it is done in classical communications~\cite{proakis_digital_2008}.
This leads to time-dependent symbol fluctuations which departs from the \gls[hyper=false]{dhd} model.
It is also claimed in~\cite{qu_high-speed_2018} that the fluctuations when measuring a symbol with \gls[hyper=false]{kkd} are smaller than in \gls[hyper=false]{dhd} because there is no halving of the signal with a balanced beamsplitter.
However, while there is no $3\unit{dB}$ loss in \gls[hyper=false]{snr} from an other input of a beamsplitter as in \gls[hyper=false]{dhd}, there is a $3\unit{dB}$ loss coming from an orthogonal spectral mode in \gls[hyper=false]{kkd} as we will discuss.
The work in Ref.~\cite{zhang_quantum_2023} also indicates that quadrature fluctuations are non-isotropic in phase-space, when measuring coherent states which is another result that departs from what is obtained with traditional balanced coherent detection schemes. 
\par In this article, we show differently that \gls[hyper=false]{kkd} is operationally similar to \gls[hyper=false]{dhd} as it can measure both quadratures of the field with a $3\unit{dB}$ \gls[hyper=false]{snr} loss compared to \gls[hyper=false]{hd}. 
We consider a model where optical and signal processing are ideal and where the electronics is noiseless and linear, so that the only noise source stems from quantum fluctuations.
Our results are obtained by considering a first-order expansion in the inverse amplitude of the local oscillator.
Unlike in~\cite{zhang_quantum_2023}, we consider normalized spectral modes instead of plane waves using the formalism of~\cite{blow_continuum_1990,fabre_modes_2020,chen_2023} as the carrier of the symbols.
In this context, the natural way to retrieve a symbol from a signal is to filter the signal with the associated spectral mode~\cite{proakis_digital_2008}.
By contrast with ~\cite{qu_high-speed_2018}, we show that, even though there is no halving of the signal such as in \gls[hyper=false]{dhd}, the \gls[hyper=false]{snr} is still halved for coherent states because of a coupling with an orthogonal spectral mode.
We stress that the proof of having the same SNR for the KKD in the quantum regime as for DHD strongly relies on the spectral degree of freedom of light, also crucial in the classical analysis of \gls[hyper=false]{kkd}. This is particularly important due to the \gls[hyper=false]{ssb} constraint~\cite{mecozzi_kramerskronig_2016,mecozzi_kramerskronig_2019}, that plays a crucial and different role in the quantum regime of KKD.
\par In addition, we employ a formulation of \gls[hyper=false]{kkd} that uses a perfectly defined Hermitian phase operator at time $t$ that is related to all the moments of the intensity operator at all time by a Hilbert transform. 
We stress that the phase operator that is defined in this manuscript - and its related first and second moments - obtained through \gls[hyper=false]{kkd} is not the phase operator obtained by a polar decomposition of the annihilation operator defined in the quantum optics literature~\cite{susskind_quantum_1964,levy-leblond_who_1976,barnett_phase_1986,pegg_unitary_1988,barnett_1989,vourdas_1993,tsang_2008}. 
Indeed, in these references, the introduced exponential of the phase operator does not commute with the number or intensity operator. 
Instead, the \gls[hyper=false]{kkd} phase operator is a different phase operator whose average value with respect to quasi-classical state matches the phase of a classical optical field.
In order to clarify the sense of the \gls[hyper=false]{kkd} phase operator, we investigate the phase that the detection reconstructs for some particular states like, \latin{e.g.}, bosonic coherent states (defined in~\cite{robertCoherentStatesApplications2021,fabre_modes_2020}), mixed states and monomode pure states. 
We emphasize that \gls[hyper=false]{kkd} allows in any case obtaining the relative phase between the known local oscillator and the quantum state of interest. 
As we shall see, the phase of the measured quantum state contains the information about both its particle-number statistics and its temporal mode structure.
\par Finally, we propose and analyze a \gls[hyper=false]{kkd} spectral tomography scheme for reconstructing the phase of the temporal wavefunction of single-photon states~\cite{blow_continuum_1990,legero_characterization_2005} from time-of-arrival-resolved detection~\cite{fabreSpectralSinglePhotons2022}. Specifically, the protocol aims to reproduce the temporally resolved intensity typically measured with a PIN photodiode—subject to the aforementioned mathematical constraints—through the measurement of the temporally resolved single-photon probability distribution using the KK relation. This method allows performing the full spectral tomography of single-photon states as it was done for instance experimentally in ~\cite{davisExperimentalSinglephotonPulse2018,davisMeasuringSinglePhotonTemporalSpectral2018,thielSinglephotonCharacterizationTwophoton2020,kurzynaVariableElectroopticShearing2022a,leiPhaseRetrievalHongOuMandel2024}.
In our proposed protocol, the single-photon state is spectrally engineered such that it can be decomposed as a coherent superposition of a large monochromatic part and a temporal structure of interest that verifies the \gls[hyper=false]{mp}~\cite{mecozzi_necessary_2016} and the \gls[hyper=false]{ssb}~\cite{mecozzi_kramerskronig_2016,mecozzi_kramerskronig_2019} conditions.
We also suggest a possible experimental implementation of the spectral engineering which relies on \gls[hyper=false]{eom}~\cite{wooten_2000,kolchin_2008,koeber_2015,kamada_2022}, optical spectral filters and a temporal-resolved single-photon detector. 
The measurement of the time-of-arrival probability single photon distribution is then used for reconstructing the phase of the single-photon state wavefunction from phase-intensity \gls[hyper=false]{kkd} relations. 
\par The article is organized as follows.
In Sec.~\ref{sec:hd-dhd} we introduce \gls[hyper=false]{hd} and \gls[hyper=false]{dhd} as examples  of coherent detections.
We derive their measured operators and the associated quantum fluctuations, as well as the \gls[hyper=false]{snr} for coherent states.
In Sec.~\ref{sec:kk}, we study \gls[hyper=false]{kkd} both in the classical regime and in the quantum regime where we derive its measured operators. 
In Sec.~\ref{sec:phase}, we investigate the phase information reconstructed by \gls[hyper=false]{kkd} for specific states such as bosonic coherent states, mixed quantum states and pure monomode states. 
In Sec.~\ref{sec:single-photon}, we present a single-photon spectral tomography scheme inspired from \gls[hyper=false]{kkd}.
Finally, in Sec.~\ref{sec:conclusion}, we summarize our results and open new perspectives.

\section{Coherent detections : homodyne detection and double homodyne detection}\label{sec:hd-dhd}
\begin{figure}
    \centering
    \includegraphics[width=\columnwidth]{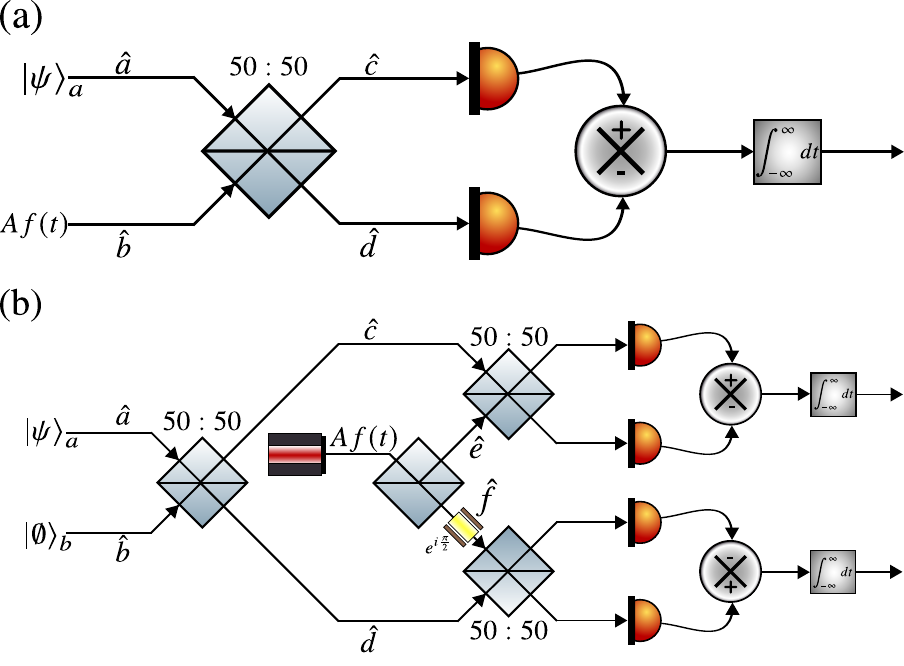}
    \caption{Setups for (a) homodyne detection and (b) double homodyne detection.
    Homodyne detection (a) allows for measuring a single quadrature of light.
    It consists in mixing the signal of interest in the port $\hat{a}$ with a strong local oscillator in the port $\hat{b}$ on a balanced beamsplitter.
    The intensities of both outputs $\hat{c}$ and $\hat{d}$ are measured with identical photodiodes.
    The difference of photocurrents is integrated and is shown to be proportional to the quadrature of the signal to measure at the phase defined by the local oscillator and in the spectral mode defined by the local oscillator.
    For the output to match the quadratures in expectation value, it has to be renormalized by measuring vacuum fluctuations.
    Double homodyne detection (b) allows for measuring both quadratures of light at the cost of an extra vacuum noise.
    The signal of interest is split in two on a balanced beamsplitter.
    Each part is measured by two identical homodyne detections with the only difference being the phase of the local oscillator.
    The first homodyne detection phase is zero to measure the position quadrature and the second one is $\pi\slash 2$ to measure the momentum quadrature.
    As a result from the initial splitting, an extra vacuum noise corresponding to the second input of the first beamsplitter is added to the measurement results.
    }
    \label{fig:setup-coherent}
\end{figure}
In typical coherent communications, the emitter sends complex symbols each associated to a spectral narrowband mode of light localized spectrally around a carrier frequency $\omega_c$ (for instance using telecom wavelength, \latin{i.e.}, $\SI{1550}{nm}$) in a spatial mode of an optical fiber.
In this context, coherent detection~\cite{kikuchi_fundamentals_2016} is a class of measurements able to measure the quadratures of the field.
The most used coherent detection schemes are \gls[hyper=false]{hd} and \gls[hyper=false]{dhd} which often serve as theoretical baselines for characterizing coherent detections.

In this section, we present the quantization of the HD and DHD using a similar formalism as in~\cite{kumar_2012,lvovsky_2016,chen_2023}, that allows us to introduce the future notation for the KKD. We will consider ideal optics, noiseless and linear electronics, meaning that the only noise taken into account is the quantum noise.
\gls[hyper=false]{hd} allows measuring a single quadrature of light in a mode specified by the \gls[hyper=false]{lo}, the electronics and the signal processing.
\gls[hyper=false]{dhd} allow measuring both quadratures of light in a mode specified by the \gls[hyper=false]{lo}, the electronics and the signal processing, at the cost of an extra $\SI{3}{dB}$ loss in \gls[hyper=false]{snr} on each quadrature compared to \gls[hyper=false]{hd} for coherent states.

\subsection{Homodyne detection}\label{sec:homodyne}

\gls[hyper=false]{hd} starts by mixing the bosonic field in spatial mode $a$ of interest with a \gls[hyper=false]{lo} in spatial mode $b$. The annihilation bosonic operator in mode $a,b$ at time $t$ are denoted $\hat{a}(t), \hat{b}(t)$. In Appendix~\ref{app:quantization}, we remind a detailed introduction of the quantization of the electromagnetic field, where the notations are more detailed. The two spatial modes are combined into a balanced beamsplitter as shown in Fig.~\ref{fig:setup-coherent}(a).
The outputs of the beam-splitter $\hat{c}(t) = (\hat{a}(t)+\hat{b}(t))/\sqrt{2}$ and $\hat{d}(t) = (\hat{a}(t) - \hat{b}(t))/\sqrt{2}$ undergo identical optical path and are measured by identical photodiodes with response function $h(t)$ and unit efficiency.
Non-unit efficiency photodiodes can be taken into account as in Ref.~\cite{banaszek_operational_1997} by formally putting a beamsplitter just before the photodiodes.
We assume the \gls[hyper=false]{lo} to be classical and replace the bosonic operator $\hat{b}(t)$ by the c-number $Af(t)$ where $A=\abs{A}e^{i\theta}$ is the amplitude and $f(t)$ is the normalized temporal mode of the \gls[hyper=false]{lo} meaning that $\int \dif t\abs{f(t)}^2=1$.
The difference of photocurrents~\cite{kumar_2012,lvovsky_2016} is written as
\begin{equation}
    \begin{aligned}
        \hat{I}_{-}(t) &= \int_{-\infty}^{\infty}\dif\tau h(t-\tau)(\hat{c}^\dag(\tau)\hat{c}(\tau) - \hat{d}^\dag(\tau)\hat{d}(\tau))\\
        &= \int_{-\infty}^{\infty}\dif\tau h(t-\tau)(A^*f^*(\tau)\hat{a}(\tau) + Af(\tau)\hat{a}^\dag(\tau)).
    \end{aligned}
\end{equation}
The output of \gls[hyper=false]{hd} is obtained by integrating $\hat{I}_-(t)$ and is proportional to the quadrature of the field of interest:
\begin{equation}
    \begin{aligned}
        \hat{o}_{\hd} &= \int_{-\infty}^\infty \dif t \hat{I}_{-}(t)\\
        &= \hat{x}_{\theta}^a\func{\tilde{\xi}}\sqrt{\int\dif\tau\abs{\xi(\tau)}^2},
    \end{aligned}
\end{equation}
where $\hat{x}_{\theta}^a\func{\tilde{\xi}}=(e^{-i\theta}\hat{a}\func{\tilde\xi}+e^{i\theta}\hat{a}^\dag\func{\tilde\xi})/2$ denotes the quadrature of the input $\hat{a}(t)$ of phase $\theta$ in the temporal mode $\tilde\xi$ (see Appendix~\ref{app:quantization}). The function $\xi(\tau)=2\abs{A}f(\tau)\int\dif t h(t-\tau)$ is defined by the modulus of the amplitude of the \gls[hyper=false]{lo}, the spectral mode of the \gls[hyper=false]{lo} and the characteristics of the photodiodes.
Hence $\tilde{\xi}(t)=\xi(t)/\sqrt{\int\dif\tau\abs{\xi(\tau)}^2}$ is the temporal mode of the measurement that also depends on the \gls[hyper=false]{lo}, the photodiodes and the electronics in general.
\par For the output to match the quadrature of the signal of interest, the output is divided by {$\sqrt{\int\dif\tau\abs{\xi(\tau)}^2}$} to be normalized.
To measure the normalization factor, the variance with the input of interest in the vacuum state $\ket{0}_{a}$ is measured:
\begin{equation}
    \Delta^2_{\ket{\vacuum}_{a}}\hat{o}_{\hd}=\frac{1}{4}\int\dif\tau\abs{\xi(\tau)}^2,
\end{equation}
and it allows renormalizing for all next measurements as
\begin{equation}
    \hat{o}_{\hd} \longleftarrow \frac{\hat{o}_{\hd}}{2\sqrt{\Delta^2_{\ket{\vacuum}_{a}}\hat{o}_{\hd}}} = \hat{x}_{\theta}^a\func{\tilde{\xi}}.
\end{equation}
\par Injecting now a coherent state defined by $\ket{\alpha} =\exp(\alpha\hat{a}^\dag\func{\chi}-\alpha^*\hat{a}\func{\chi})\ket{\vacuum}$, with $\alpha\in\mathbb{C}$, yields
\begin{equation}
    \begin{aligned}
        \bra{\alpha}\hat{o}_{\hd}\ket{\alpha}  &= \Re{e^{i\theta}\alpha\inner{\tilde{\xi}}{\chi}},\\
        \Delta^2_{\ket{\alpha}}\hat{o}_{\hd} &= \frac{1}{4},
    \end{aligned}
\end{equation}
where we have defined the scalar product $\inner{\tilde{\xi}}{\chi}=\int dt \tilde{\xi}^*(t)\chi(t)$ which is the overlap between the mode of the coherent state $\chi$ and the mode of the overall detection $\tilde{\xi}$. Therefore, the expectation value of $\hat{o}_{\hd}$  is the quadrature of angle $\theta$ of the coherent states if the spectral mode of the state matches the spectral mode of the \gls[hyper=false]{hd}.
Then, the \gls[hyper=false]{snr} becomes
\begin{equation}
    \snr_{\hd} = \frac{\bra{\alpha}{\hat{o}_{\hd}}\ket{\alpha}}{\Delta^2_{\ket{\alpha}}\hat{o}_{\hd}} = 4\Re{e^{-i\theta}\alpha}.
\end{equation}

\subsection{Double homodyne detection}\label{sec:double-homodyne}
While \gls[hyper=false]{hd} measures a single quadrature of light, \gls[hyper=false]{dhd}  allows measuring the two orthogonal quadratures of light with a setup as shown in Fig.~\ref{fig:setup-coherent}(b).
It consists in splitting the signal of interest in two with a first balanced beamsplitter, where the outputs spatial annihilation operators at time $t$ are given by: $\hat{c}(t) = \frac{1}{\sqrt{2}}(\hat{a}(t) + \hat{b}(t)),
        \hat{d}(t) = \frac{1}{\sqrt{2}}(\hat{a}(t) - \hat{b}(t))$. 
Now, $\hat{b}(t)$ denotes the second input of the first beamsplitter of the splitted signal of interest.
The outputs $c,d$ are then detected with two \gls[hyper=false]{hd}s, one with an angle $\theta = 0$ to measure the position quadrature $\hat{q}$ and one with an angle $\theta = \pi/2$ to measure the momentum quadrature $\hat{p}$.
The \gls[hyper=false]{lo}s used for the two \gls[hyper=false]{hd}s, in spatial modes $e,f$ are taken in the classical limit and are replaced by c-numbers, $\abs{A}f(t)/\sqrt{2}$ and $\abs{A}e^{i\pi/2}f(t)/\sqrt{2}$.
The $\sqrt{2}$ factor comes from the fact that the same laser is split into two to generate both \gls[hyper=false]{lo}s. Using the results of Sec.~\ref{sec:homodyne} the two unnormalized outputs are written as
\begin{equation}
    \begin{aligned}
        \hat{o}_{\dhd}^{(q)} &= \left(\hat{q}_{a}\func{\tilde{\xi}} + \hat{q}_b\func{\tilde{\xi}}\right)\sqrt{\int\dif\tau\abs{\xi(t)}^2},\\
        \hat{o}_{\dhd}^{(p)} &= \left(\hat{p}_{a}\func{\tilde{\xi}} - \hat{p}_b\func{\tilde{\xi}}\right)\sqrt{\int\dif\tau\abs{\xi(t)}^2},\\
    \end{aligned}
\end{equation}
where $\xi(\tau) = \abs{A}f(\tau)\int\dif th(t-\tau)$ is defined similarly as in \gls[hyper=false]{hd} by the amplitude of the \gls[hyper=false]{lo}, the spectral mode of the \gls[hyper=false]{lo} and the characteristics of the photodiodes, and $\tilde{\xi}(t)$ is the normalized mode associated to $\xi(t)$.
Again, as in \gls[hyper=false]{hd}, the outputs must be renormalized a posteriori so that they match the quadratures of the signal of interest in expectation value.
When the input $\hat{a}(t)$ is kept empty, the variances on the measured operators are
\begin{equation}
    \begin{aligned}
        \Delta_{\ket{\vacuum}_{ab}}^2\hat{o}_{\dhd}^q &= \frac{1}{2}\int\dif\tau\abs{\xi(\tau)}^2,\\
        \Delta_{\ket{\vacuum}_{ab}}^2\hat{o}_{\dhd}^p &= \frac{1}{2}\int\dif\tau\abs{\xi(\tau)}^2,
    \end{aligned}
\end{equation}
which allows us to renormalize the outputs as follows:
\begin{equation}
    \begin{aligned}
        \hat{o}_{\dhd}^q &\longleftarrow \frac{\hat{o}_{\dhd}^q}{\sqrt{2}\sqrt{\Delta_{\ket{\vacuum}_{ab}}^2\hat{o}_{\dhd}^q}} = \hat{q}_{a}\func{\tilde{\xi}}+\hat{q}_{b}\func{\tilde{\xi}},\\
        \hat{o}_{\dhd}^p &\longleftarrow \frac{\hat{o}_{\dhd}^p}{\sqrt{2}\sqrt{\Delta_{\ket{\vacuum}_{ab}}^2\hat{o}_{\dhd}^p}} = \hat{p}_{a}\func{\tilde{\xi}}-\hat{p}_{b}\func{\tilde{\xi}}.
    \end{aligned}
\end{equation}
In this form, the outputs consist of the quadratures to be measured and additional terms corresponding to the quadratures of the second input of the first beamsplitter, referred to as vacuum noise.
\par If the input state is now a coherent state $\ket{\alpha}$ in the same spectral mode as the one of the detector, the moments of the outputs of \gls[hyper=false]{dhd} are
\begin{equation}
    \begin{aligned}
        \bra{\alpha}{\hat{o}_{\dhd}^q}\ket{\alpha} &= \Re{\alpha}, &\bra{\alpha}{\hat{o}_{\dhd}^p}\ket{\alpha} &= \Im{\alpha},\\
        \Delta^2_{\ket{\alpha}}\hat{o}_{\dhd}^q &= \frac{1}{2}, &\Delta^2_{\ket{\alpha}}\hat{o}_{\dhd}^p &= \frac{1}{2},
    \end{aligned}
\end{equation}
and the \gls[hyper=false]{snr} for a single quadrature is then
\begin{equation}
    \begin{aligned}
        \snr_{\dhd}^q &= 2\Re{\alpha}, &\snr_{\dhd}^p&=2\Im{\alpha},
    \end{aligned}
\end{equation}
which is half that of \gls[hyper=false]{hd}.
This result is the famous \gls[hyper=false]{dhd} $\SI{3}{dB}$ loss for the \gls[hyper=false]{snr} compared to \gls[hyper=false]{hd}.

\section{Kramers-Kronig detection}\label{sec:kk}
\begin{figure}
    \centering
    \includegraphics[width=\columnwidth]{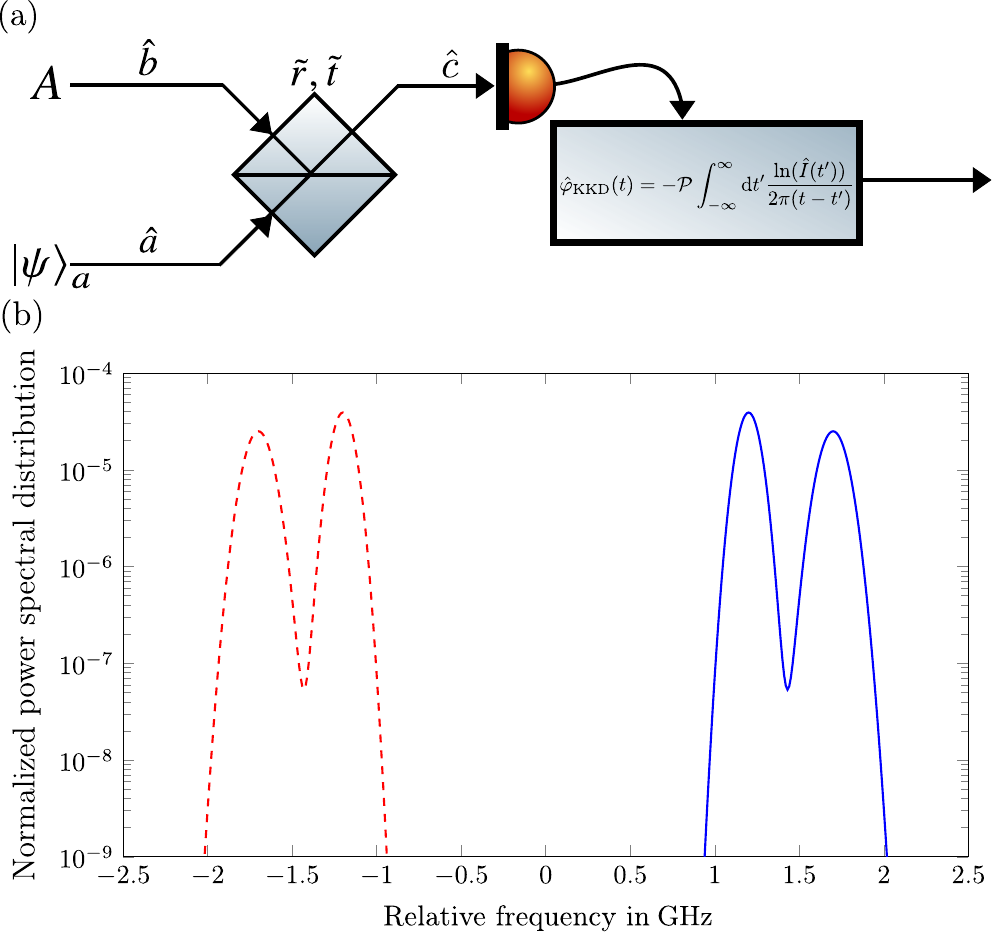}
    \caption{(a) Setup for Kramers-Kronig detection and (b) example of single sideband signal.
    For Kramers-Kronig detection (a), the signal of interest is injected in the $\hat{a}$ input of a \textit{highly unbalanced} beamsplitter while a monochromatic strong and classical local oscillator is injected in the $\hat{b}$ input.
    The signal of interest verifies the single sideband constraint.
    A single output $\hat{c}$ intensity is measured by a photodiode and the intensity at each instant is recorded.
    The phase of the output $\hat{c}$ is reconstructed by taking the Hilbert transform of the logarithm of the measured intensity.
    Since the local oscillator is known, it is possible to reconstruct the input of interest $\hat{a}$ with the reconstructed $\hat{c}$ output.
    Single sideband signals (b) are signals spectrally on the right of the carrier frequency represented here by the zero value.
    The plain (blue) signal is an example of the spectrum of a single sideband mode (logarithmic scale), whereas the dashed (red) signal is the complex conjugate of the blue signal and is not a single sideband mode as it lies on the left of the carrier frequency.
    For Kramers-Kronig this conjugated spectral mode is a source of extra noise.
    }
    \label{fig:kk_setup}
\end{figure}

\gls[hyper=false]{kkd} is a direct detection relying on signal engineering to recover the phase of the measured signal~\cite{mecozzi_kramerskronig_2016,mecozzi_kramerskronig_2019}.
A setup for \gls[hyper=false]{kkd} is presented in Fig.~\ref{fig:kk_setup}(a).
It consists in mixing a signal of interest and a monochromatic reference in the form of a \gls[hyper=false]{lo} on an unbalanced beamsplitter and measuring a single output.
It has been shown that under \gls[hyper=false]{mp} constraint and \gls[hyper=false]{ssb} constraint, it is possible to reconstruct the phase of $\hat{c}(t)$ from its intensity through a Hilbert transform.
\par In this section, we present the results in the classical regime and the necessary constraints for \gls[hyper=false]{kkd} to work.
Then, we present our results in the quantum regime and show that \gls[hyper=false]{kkd} allows reconstructing both quantum quadratures of the field at the cost of an extra vacuum noise coming from an independent spectral mode.

\subsection{Kramers-Kronig detection in the classical regime}
In the classical regime, for perfoming the phase reconstruction using the KK relation, the \gls[hyper=false]{mp} constraints must be verified. 
The \gls[hyper=false]{lo} is monochromatic with no phase at the carrier frequency, meaning that its amplitude can be written as $b(t) = A$.
The amplitude $A$ is assumed to be real and defines thus a phase reference.
In order to satisfy the \gls[hyper=false]{mp} constraint, the \gls[hyper=false]{lo} contribution has to be larger than the signal to be measured~\cite{mecozzi_necessary_2016}:
\begin{equation}
    \tilde{r}A>\tilde{t}\abs{a(t)}.\label{eq:mp}
\end{equation}
Additionally, the input of interest must verify the \gls[hyper=false]{ssb} constraint which consists in $a(t)$ being spectrally on the right of the \gls[hyper=false]{lo}, \latin{i.e.} 
\begin{equation}
    \fourier(a)(\omega\leq 0) = 0,\label{eq:ssb}
\end{equation}
where $\fourier(f)(\omega) = \int_{-\infty}^\infty\dif t f(t)e^{i\omega t}$ denotes the Fourier transform with $\omega$ the relative frequency with respect to the carrier frequency.
When the \gls[hyper=false]{mp} constraint is verified, the logarithm of the output $c(t)=\tilde{r}A + \tilde{t}a(t)$, where $\tilde{r}$ and $\tilde{t}$ are the reflection and transmission coefficients of the beamsplitter, can be expressed as a power series since $\tilde{r}A>\tilde{t}\abs{a(t)}$:
\begin{equation}
    \begin{aligned}
        \ln(c(t)) &= \ln(\tilde{r}A + \tilde{t}a(t))\\
        &= \ln(\tilde{r}A) + \sum_{n=0}^\infty (-1)^n\frac{\tilde{t}^na(t)^n}{\tilde{r}^nA^nn}.
    \end{aligned}
\end{equation}
The total term is also \gls[hyper=false]{ssb} since convolutions in frequency of \gls[hyper=false]{ssb} functions are \gls[hyper=false]{ssb} functions.
On the other hand the logarithm of the $c(t)$ output can be expressed in terms of the phase $\argument(c(t))$ and the intensity $I(t) = \abs{c(t)}^2$ as follows
\begin{equation}
    \ln(c(t)) = \frac{1}{2}\ln(I(t)) + i\argument(c(t)).
\end{equation}
Since the total expression is \gls[hyper=false]{ssb}, it verifies traditional Kramers-Kronig relations for electrical susceptibilities~\cite{kramers_1927} where the time and frequency have been swapped.
The intensity being measurable, it is possible to reconstruct the phase from the time measurement results~\cite{mecozzi_kramerskronig_2016,mecozzi_kramerskronig_2019} with a Hilbert transform~\cite{king_2009}, i.e.:
\begin{equation}
    \varphi_{\kk}(t) = -\pval\int_{-\infty}^\infty \dif t' \frac{\ln(I(t'))}{2\pi(t-t')},\label{eq:hilbert-transform}
\end{equation}
where $\varphi_{\kk} = \argument(c(t))$ in classical regime and where $\pval$ denotes Cauchy principal value~\cite{king_2009}.
The output $c(t)$ is reconstructed as follows: $c(t) = \sqrt{I(t)}e^{i\varphi_{\kk}(t)}$ and the input signal reads
\begin{equation}
    a(t) = \frac{c(t)-\tilde{r}A}{\tilde{t}} = \frac{\sqrt{I(t)}e^{i\varphi_{\kk}(t)} - \tilde{r}A}{\tilde{t}}.
\end{equation}
The complex symbols and the quadratures for a given spectral mode $f(t)$ are finally recovered with a projection integral:
\begin{equation}
    \begin{aligned}
        a\func{f} &= \int_{-\infty}^\infty\dif t f^*(t) a(t),\\
        q\func{f} &= \frac{a\func{f}+a^*\func{f}}{2},\\
        p\func{f} &= \frac{a\func{f}-a^*\func{f}}{2i}
    \end{aligned}
\end{equation}
where $f(t)$ is a \gls[hyper=false]{ssb} mode.
Indeed $a(t)$ is supposed to be only composed of \gls[hyper=false]{ssb} modes to verify the \gls[hyper=false]{ssb} constraint.
Since both quadratures can be recovered, \gls[hyper=false]{kkd} is indeed a direct detection acting as a coherent detection measuring both quadratures in the classical regime.

\subsection{Kramers-Kronig detection in the quantum regime}\label{sec:kk_quantum}

The quantization of the inputs $a(t)$ and $b(t)$ and the output $c(t)$ leads to their replacement as bosonic quantum operators $\hat{a}(t)$, $\hat{b}(t)$ and $\hat{c}(t)$, similarly to what it was done in Sec.~\ref{sec:homodyne}.
\par Comparatively to traditional balanced detections only a single output of the beamsplitter is measured for \gls[hyper=false]{kkd} and thus the second output is formally traced out.
In the case of coherent states at both inputs this tracing operation does not modify the state of the measured output since the output state is a product of two coherent states.
However for a general state at the input, the tracing out operation will introduce a supplementary noise. To circumvent this, we propose to use a highly unbalanced beamsplitter to mitigate the effect of the trace on the second output.
The beamsplitter must satisfy the condition \(\tilde{t} \to 1\), ensuring that the fluctuations in the non-measured output do not impact the signal of interest. The reflection coefficient must then be $\tilde{r} \to 0$ and thus the local oscillator amplitude must be chosen accordingly.
\par We treat the case of a classical \gls[hyper=false]{lo} meaning that $\hat{b}(t)$ is replaced by a c-number $A$, since the LO is still real and monochromatic.
For practical purposes, however, we extend the \gls[hyper=false]{mp} constraints to an asymptotic form, which allows us to perform a Taylor expansion of the logarithm in the reconstructed phase, defined in Eq.~(\ref{eq:hilbert-transform}):
\begin{equation}
    \tilde{r}A \gg \tilde{t}\abs{\expval{a(t)}},
\end{equation}
where the expectation value is taken on the unknown state of interest.
This asymptotic constraint is verified in practice for traditional coherent detections in quantum optics protocols since the \gls[hyper=false]{lo} is supposed to be classical and the quantum signal to measure is supposed to be of only few photons.
The intensity of a single output, $\hat{I}(t) = \hat{c}^\dag(t)\hat{c}(t)$, is measured. 
Performing the Hilbert transform on the logarithm of the intensity, defines a \gls[hyper=false]{kkd} phase operator written as
\begin{equation}
    \hat{\varphi}_{\kk}(t) = -\pval\int_{-\infty}^\infty\dif t' \frac{\ln(\hat{I}(t'))}{2\pi(t-t')}.\label{eq:phase-op}
\end{equation}
This phase operator is well-defined, Hermitian and commutes with the intensity, making it fundamentally different from the phase operator of the quantum optics literature which is not Hermitian and does not commute with the number operator~\cite{susskind_quantum_1964,levy-leblond_who_1976,barnett_phase_1986,pegg_unitary_1988,barnett_1989,tsang_2008} (see Appendix~\ref{app:phase} for a detailed description).

Taking into account the strong classical monochromatic \gls[hyper=false]{lo}, and thus that $\hat{I}(t)\approx \tilde{r}^2A^2$ at first order, the logarithm is Taylor expanded.
The phase $\hat{\varphi}_{\kk}(t)$ is then written as a function of the quadratures operators associated to all plane waves of positive frequency relative to the carrier:
\begin{align}
    \hat{\varphi}_{\kk}(t) = & \int_{0}^\infty \frac{\dif\omega}{2\pi} \frac{\tilde{t}}{\tilde{r}} \frac{1}{A} \left[ (\hat{p}(\omega)-\hat{p}(-\omega)) \cos(\omega t) \right. \nonumber \\
    & \left. - (\hat{q}(\omega)+\hat{q}(-\omega)) \sin(\omega t) \right] \nonumber \\
    & + O\left( \frac{1}{A^2} \right),
\end{align}
where $O(1\slash A^2)$ denotes an operator of expectation and cumulants of order $O(1\slash A^2)$ (see Appendix~\ref{app:kk-dsp} for a detailed derivation of this result).

\par We define the polar decomposition of the annihilation operator at time $t$ from the intensity and phase operator that are measured as $\hat{c}_{\kk}(t) = \sqrt{\hat{I}(t)}e^{i\hat{\varphi}_{\kk}(t)}$, inspired by the polar decomposition of the phase operator in a given mode. 
The reconstructed input of interest, at first order of the LO intensity can thus be written as
\begin{equation}
    \begin{aligned}
        \hat{a}_{\kk}(t) &= \frac{\hat{c}_{\kk}(t) - \tilde{r}A}{\tilde{t}}\\
        &=\int_0^\infty \frac{\dif \omega}{2\pi}(\hat{a}(\omega) + \hat{a}^\dag(-\omega))e^{i\omega t} + O\left(\frac{1}{A}\right).\label{eq:a_kk}
    \end{aligned}
\end{equation}
The signal components at negative relative frequency appear in the reconstructed signal, even though the emitted signal is supposed to be \gls[hyper=false]{ssb}.
This contribution corresponds to vacuum contribution and will have an impact on the calculation of fluctuations.
The quadratures in a given \gls[hyper=false]{ssb} spectral mode $f(t)$ can be reconstructed with a projection integral yielding the final result:
\begin{equation}
    \begin{aligned}
        \hat{o}^q_{\kk}\func{f} &= \hat{q}\func{f} + \hat{q}\func{f^*}+O\left(\frac{1}{A}\right),\\
        \hat{o}^p_{\kk}\func{f} &= \hat{p}\func{f} - \hat{p}\func{f^*}+O\left(\frac{1}{A}\right),\label{eq:quadrature-kk}\\
    \end{aligned}
\end{equation}
where again, the extra quadrature operators associated to the mode $f^*(t)$ correspond to vacuum fluctuations.
Indeed, in our convention, the mode $f^*(t)$ correspond to a spectral mode of negative frequency relative to the carrier frequency.
An example is shown in Fig.~\ref{fig:kk_setup}(b), where the plain (blue) spectrum corresponds to a \gls[hyper=false]{ssb} mode $f(t)$ and the dashed (red) spectrum corresponds to the complex conjugate $f^*(t)$.
This shows that \gls[hyper=false]{kkd} is equivalent to \gls[hyper=false]{dhd} since both measure the two orthogonal quadratures of light with an extra vacuum noise penalty.
However, the extra vacuum noise comes from an empty independent spatial mode for \gls[hyper=false]{dhd} whereas it comes from an empty independent spectral mode for \gls[hyper=false]{kkd}.

\subsection{Discussion regarding experimental implementation}\label{sec:kk_implem}

\begin{figure}
    \centering
    \includegraphics[width=\columnwidth]{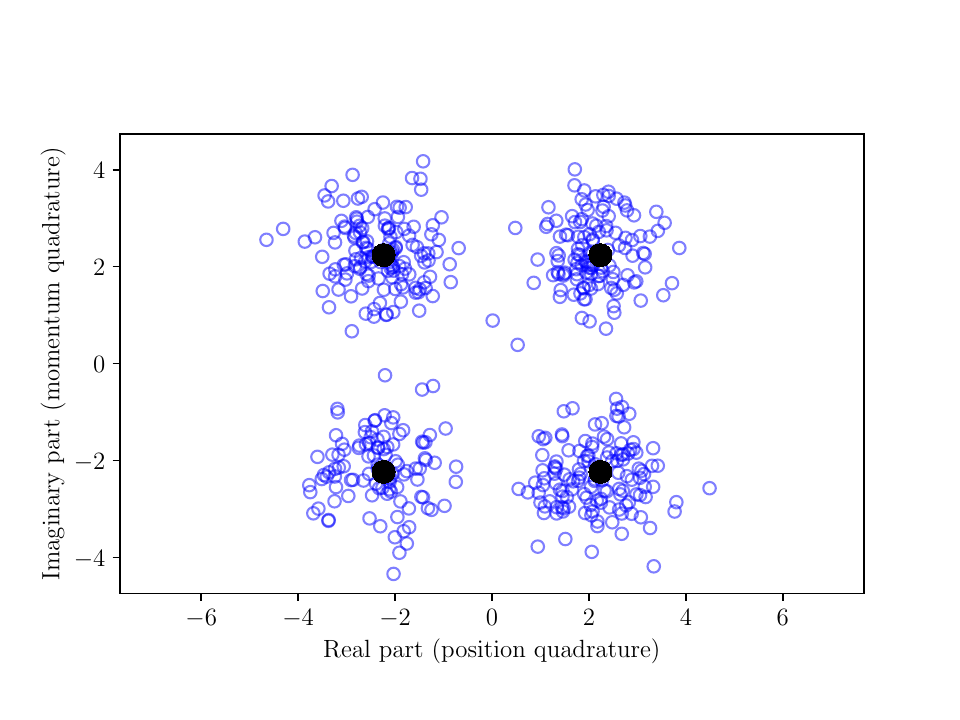}
    \caption{Simulation of a symbol reconstruction process with Kramers-Kronig detection (phasor diagram). 
    The empty circles markers (blue) denote the measured symbols whereas the full circles (black) denote the emitted symbols.
    The emitted symbols are coherent states belonging to a 4-QAM constellation and are each associated to a time shifted Gaussian pulse of bandwidth $22$ GHz.
    Each pulse is composed of 10 photons in average.
    The simulated link is a back-to-back link where the emitter directly sends into the receiver.
    Finally, the quantum noise is generated at the photodiode where the output photocurrent follows a Normal law with variance that of $\hat{I}(t)$ when the local oscillator is taken classical and completely deterministic.
    }
    \label{fig:4QAM}
\end{figure}
In the case of a practical implementation several things have to be taken into account.
First, in order to perform the substraction of Eq.~(\ref{eq:a_kk}) it is necessary to know the local oscillator amplitude $A$ beforehand. 
The phase of $A$ serves as a phase reference since it is assumed to be real.
The modulus of $A$ must be measured before performing the \gls[hyper=false]{kkd} coherent detection scheme. Secondly, in Sec.~\ref{sec:kk_quantum} the quantities are written in normalized unit such that the variances of the quadratures for vacuum gives $1\slash2$.
The measurements must then be renormalized for the outputs to match the quadratures in expectation value.
The process is the same as that for \gls[hyper=false]{dhd}: the variance for vacuum is measured and the next measurements are renormalized by the vacuum variance which gives
\begin{equation}
    \begin{aligned}
        \hat{o}_{\kk}^q &\longleftarrow \frac{\hat{o}_{\kk}^q}{\sqrt{2}\sqrt{\Delta_{\ket{\vacuum}_{a}}^q\hat{o}_{\kk}^q}},
        &\hat{o}_{\kk}^p &\longleftarrow \frac{\hat{o}_{\kk}^p}{\sqrt{2}\sqrt{\Delta_{\ket{\vacuum}_{a}}^p\hat{o}_{\kk}^p}}.
    \end{aligned}
\end{equation}
\par In Fig.~\ref{fig:4QAM}, we show a simulation of a symbol reconstruction with \gls[hyper=false]{kkd}.
In this simulation the simulated state to be measured is a classical mixture of random coherent states belonging to a 4-QAM~\cite{proakis_digital_2008} constellation with an average of $10$ photons per symbols.
The temporal mode for each symbol is a time-shifted Gaussian.
The local oscillator is a known classical monochromatic signal and has an intensity of $10^{5}\mathrm{photon.s^{-1}}$.
The source for quantum noise is generated at the photodiode.
The output current follows a Gaussian law of variance corresponding to the variance of $\hat{I}(t)$ when the \gls[hyper=false]{lo} is completely classical and deterministic. 

\par Finally, compared to traditional balanced detections, since \gls[hyper=false]{kkd} is a direct detection scheme, it could be argued that it is robust to the phase noise~\cite{quContinuousVariableQuantumKey2018} which is a major issue for traditional coherent detection scheme.
However, we have shown that the \gls[hyper=false]{lo} still acts as a phase reference and therefore \gls[hyper=false]{kkd} should need phase stabilization of the \gls[hyper=false]{lo} in the case when it is not sent by the emitter.
The impact of phase noise should still be different for \gls[hyper=false]{kkd} than for regular balanced detections and could be investigated.

\section{Kramers-Kronig phase operator}\label{sec:phase}
The phase operator defined via \gls[hyper=false]{kkd} signal processing is different from the phase operator of the quantum optics literature.
Indeed, \gls[hyper=false]{kkd} phase operator is Hermitian, and commutes with the intensity operator both at time $t$ (and not projected into a given mode). Whereas the quantum optics phase operator is not Hermitian and does not commute with the intensity. Besides, both phase and intensity operators are defined into a given temporal mode ~\cite{susskind_quantum_1964,levy-leblond_who_1976,barnett_phase_1986,pegg_unitary_1988,barnett_1989,tsang_2008}.
In this section, we investigate the phase information measured by \gls[hyper=false]{kkd} detection for some states of light such as bosonic coherent states, mixed states as well as monomode pure states.

\subsection{Bosonic coherent states}\label{sec:kk-bosonic}
We start with a bosonic coherent state defined as
\begin{equation}
    \begin{aligned}
        \cohtket{\psi} &= \exp{\int_{-\infty}^\infty\dif t \psi(t)\hat{c}^\dag(t) - \hc}\ket{\vacuum}\\
        &= \exp{\sqrt{\left(\int\dif t \abs{\psi(t)}^2\right)}\hat{c}^\dag\func{\frac{\psi}{\sqrt{\int\dif t \abs{\psi(t)}^2}}} - \hc}\ket{\vacuum}.
    \end{aligned}
\end{equation}
The bosonic coherent  state~\cite{PhysRevLett.88.027902,robertCoherentStatesApplications2021,fabre_modes_2020} is intrisically monomodal as it is possible to normalize the temporal distribution of the bosonic coherent state as it is done in the second equality.
We assume $\psi(t) = A + \chi(t)$ where $\chi(t)$ is \gls[hyper=false]{ssb} and $A$ is real and much larger than $\chi(t)$.
Under these conditions the phase of $\psi(t)$ is the Hilbert transform of the logarithm of its modulus.
Such a state corresponds to the state of the measured output in Fig.~\ref{fig:kk_setup}(a) where the local oscillator state is a strong monochromatic coherent state, the input of interest state is a coherent state in a \gls[hyper=false]{ssb} mode and the transmittivity of the beamsplitter is close to one~\cite{paris_1996a}.
The characteristics of the beamsplitter have been taken into account into $A$ and $\chi(t)$ by rescaling and with the assumption that $A\gg\chi(t)$ for all $t$.
\par In order to calculate the moments of the intensity operator, one has to take into account the time response $h(t)\approx\delta(t)$ (or the bandwidth) of the photodiode, \latin{i.e.}: $\hat{I}(t) = \int_{-\infty}^\infty\dif\tau h(t-\tau)\hat{c}^\dag(\tau)\hat{c}(\tau)$.
The response function should be close to a Dirac function compared to $\psi(t)$ at the first order.
We obtain then
\begin{equation}
    \begin{aligned}
        \cohtbra{\psi}{\hat{I}(t)}\cohtket{\psi} &= \int\dif\tau h(t-\tau)\abs{\psi(\tau)}^2 \approx \abs{\psi(t)}^2,\\
        \Delta^2_{\cohtket{\psi}}\hat{I}(t) &= \int\dif\tau\abs{h(t-\tau)}^2\abs{\psi(\tau)}^2.
    \end{aligned}
\end{equation}
The variance of the intensity for coherent bosonic states depends on the expectation value similarily as the number operator for regular coherent states. 
However, it also depends on the time response of the photodiode.
With the results of Appendix~\ref{app:kk-dsp}, we calculate the first moment of the corresponding \gls[hyper=false]{kkd} phase operator:
\begin{equation}
    \begin{aligned}
        \cohtbra{\psi}{\hat{\varphi}_{\kk}(t)}\cohtket{\psi} &= -\pval\int_{-\infty}^\infty\dif t' \frac{\expval{\hat{I}(t')}-A^2}{A^2 2\pi(t-t')} + O\left(\frac{1}{A^2}\right)\\
        &= -\pval\int_{-\infty}^\infty\dif t'\frac{\ln{\abs{\psi(t')}^2}}{2\pi(t-t')} + O\left(\frac{1}{A^2}\right)\\
        &= \argument{\psi(t)} + O\left(\frac{1}{A^2}\right),
    \end{aligned}\label{eq:phasecoht}
\end{equation}
which matches the phase of the bosonic coherent state instantaneous amplitude.
We remind that this phase is the relative phase since we have assumed the strong classical monochromatic \gls[hyper=false]{lo} to be real, setting the reference.
Using the temporal response of the photodiode, the variance of the \gls[hyper=false]{kkd} phase operator can be written as
\begin{align}
    \Delta^2\hat{\varphi}_{\kk}(t) &= \pval \int_{-\infty}^{\infty} \dif t' \, \dif t'' \, \dif\tau \, \frac{h(t'-\tau)h(t''-\tau) \abs{\psi(\tau)}^2}{\pi^2 A^4 (t-t')(t-t'')} \notag \\
    &\quad + O\left(\frac{1}{A^3}\right) \notag \\
    &= O\left(\frac{1}{A^2}\right).
\end{align}
where for the second equality, we have used the fact that $\abs{\psi(t)}^2$ is of order $A^2$ and $h(t)$ is of order one.
Hence, the stronger the monochromatic reference is, the smaller the fluctuations of \gls[hyper=false]{kkd} the measured phase is.
This result is consistent as the phase of the field in the classical regime matches the \gls[hyper=false]{kkd} phase of strong bosonic coherent states.

\subsection{Mixed states}\label{ref:mixed-kk}
We consider a mixture of bosonic coherent states
\begin{equation}
    \hat{\rho} = \int\Dif^2\func{\psi} \pfunc\func{\psi}\cohtket{\psi}\cohtbra{\psi},
\end{equation}
where $\int\Dif\func{\psi} = \int\int\Dif\func{\Re{\psi}}\Dif\func{\Im{\psi}}$ denotes a functional integral and $\pfunc\func{\psi}$ denotes a P-functional distribution~\cite{glauber_1963,sudarshan_1963} which contains the information on both the photon-number statistics and the spectral degree of freedom.
For example, for a pure bosonic coherent state $\cohtket{\beta}$, $\pfunc\func{\alpha}$ is a Dirac functional distribution, \latin{i.e.}, $\pfunc\func{\alpha}=\delta(\func{\alpha}-\func{\beta})$. 
\par The measurement of the exact expectation value of \gls[hyper=false]{kkd} phase operator necessitates to measure all the moments of the intensity operator due to the high non-linearity of the logarithm (see Eq.~(\ref{eq:phase-op})).
In Appendix~\ref{app:mixedstate}, we show the expression of the first moments of the intensity operator for calculating the average value of the phase operator.
We assume the state $\hat{\rho}$ to be a mixture of bosonic coherent states, all with the same monochromatic \gls[hyper=false]{lo} of amplitude $A$ and a \gls[hyper=false]{ssb} signal with a small amplitude compared to the \gls[hyper=false]{lo} as in Sec.~\ref{sec:kk-bosonic}.
This state corresponds to the output of Fig.~\ref{fig:kk_setup}(a) when the \gls[hyper=false]{lo} is a pure strong monochromatic coherent state and the input $\hat{a}$ of interest is in a mixed state of \gls[hyper=false]{ssb} coherent state and the beamsplitter transmission coefficient is close to one~\cite{paris_1996a}.
The exact average value of the phase operator can be cast as $ \expval{\hat{\varphi}_{\kk}(t)}=\tr(\hat{\rho} \hat{\varphi}_{\kk}(t))$:
\begin{multline}
    \expval{\hat{\varphi}_{\kk}(t)} = 
    -\pval \int \Dif \func{\psi} \dif t' \sum_{1 \leq l \leq k \leq n}^\infty 
    \frac{\pfunc\func{\psi}}{2\pi(t-t')}
    \frac{(-1)^{n+k+1}}{n} \\
    \times \binom{n}{k}  \stirling{k}{l} 
    \frac{\abs{\psi(t')}^{2l}}{A^{2k}}
\end{multline}
where $\stirling{n}{k} = \frac{1}{k!} \sum_{j=0}^{k} (-1)^{k-j} \binom{k}{j} j^{n}$ are the Stirling numbers of the second kind~\cite{mansour_2015}.
\par In the case of a pure bosonic coherent state $\hat{\rho}=\cohtket{\psi}\cohtbra{\psi}$ with a \gls[hyper=false]{lo} as well as a \gls[hyper=false]{ssb} signal in the same fashion as in Sec.~\ref{sec:kk-bosonic}, the expectation value of \gls[hyper=false]{kkd} phase can be exactly written:
\begin{multline}
    \expval{\hat\varphi_{\kk}(t)} =\\
    -\pval\int\sum_{1 \leq l \leq k \leq n}^\infty\frac{\dif t'}{2\pi(t-t')}\frac{(-1)^{n+k+1}}{n}\binom{n}{k}\stirling{k}{l}\frac{\abs{\psi(t')}^{2l}}{A^{2k}}
\end{multline}
which is an equivalent expression of the phase of the bosonic coherent state as developed in Eq.~(\ref{eq:phasecoht}) at first order. 
\par Now, in the case of an incoherent sum of bosonic coherent states $\hat{\rho}=\frac{1}{2}(\cohtket{\alpha}\cohtbra{\alpha}+\cohtket{\beta}\cohtbra{\beta})$, keeping the hypothesis that both amplitudes of the bosonic coherent states are much smaller than the amplitude of the \gls[hyper=false]{lo}, the measured phase will be the sum of the relative phases between the local oscillator and each of the two \gls[hyper=false]{ssb} signals. 
For the most general mixed state, $\hat{\rho}=\sum_{i} p_{i} \ket{[\psi]_{i}}\bra{[\psi]_{i}}$, which corresponds to the $\pfunc$-functional distribution $\pfunc\func{\psi}=\sum_{i} p_{i} \delta(\func{\psi}-\func{\psi_i})$, the mean value of the measured phase would be the convex sum of all components of the mixture.

\subsection{Pure monomode state}\label{sec:kk-monomode}
In this section, unlike the precedent sections, we consider the case of a \gls[hyper=false]{lo} with a non-zero phase to show that \gls[hyper=false]{kkd} determines the relative phase between the known local oscillator and the state of interest.
This relative measured phase is composed of the particle-number statistics and the temporal mode structure of the state as we shall see.
\par We denote the state to be measured in the input $\hat{a}$ of the setup in Fig.~\ref{fig:kk_setup}(a) as
\begin{equation}
    \ket{\psi}_{a} = \sum_{n=0}^{\infty} c_{n}\ket{n}_{f} = \sum_{n=0}^{\infty} c_{n} \frac{(\hat{a}^\dag\func{f})^{n}}{\sqrt{n!}} \ket{\vacuum},
\end{equation}
where $\hat{a}^\dag\func{f}$ is the creation operator associated to the mode $f$ (see Appendix~\ref{app:quantization} for a precise definition) which is normalized, \latin{i.e.}: $\int_{-\infty}^\infty\dif t \abs{f(t)}^2=1$.
By expressing the state in this manner, we factorize it into a product of a temporal component, represented by the mode function $f(t)$, and a particle-number degree of freedom, represented by the coefficients $c_{n}$.
This factorization assumes the separability between the temporal mode $f(t)$ and the particle-number degree of freedom $c_{n}$, allowing us to treat them independently in the analysis.
The direct detection in the output spatial port $\hat{c}$ and in the temporal mode $h$ can thus be expressed as
\begin{equation}
    \begin{aligned}
      \hat{c}^\dag\func{h}\hat{c}\func{h}& = \tilde{t}^{2}\hat{a}^\dag\func{h}\hat{a}\func{h} + \tilde{r}^{2}\hat{b}^\dag\func{h}\hat{b}\func{h}\\&\quad + \tilde{t}\tilde{r}\big(\hat{a}^\dag\func{h}\hat{b}\func{h} + \hat{b}^\dag\func{h}\hat{a}\func{h}\big).  
    \end{aligned}
\end{equation}
\par For an instantaneous measurement at time $t$, matching the one used for the photodiodes, we approximate the mode function $h(t')$ by $\delta(t-t')$. 
While the Dirac delta function $\delta(t-t')$ is not physically realizable as it cannot define a mode due to its lack of normalization, we use this approximation assuming that the detector’s response time is much shorter than the temporal wavepacket duration of the fields in spatial modes $\hat{a}$ and $\hat{b}$. 
The input of the beamsplitter consists of the wavefunction of the state of interest in spatial mode $\hat{a}$ and a coherent state in spatial mode $\hat{b}$, used as a \gls[hyper=false]{lo}. 
This combined input state can be written as:
\begin{equation}
   \ket{\psi}_{a}\ket{A_g}_{b} =\left( \sum_{n=0}^{\infty} c_{n} \frac{(\hat{a}^\dag\func{f})^{n}}{\sqrt{n!}}\right) \left( \sum_{n=0}^{\infty} e^{-\frac{|A|^2}{2}}\frac{A^{n}}{\sqrt{n!}} \frac{(\hat{b}^\dag\func{g})^{n}}{\sqrt{n!}}\right)\ket{\vacuum}, 
\end{equation}
where $f$ is the spectral mode associated to $\hat{a}$, $g$ is the spectral mode associated to $\hat{b}$ and $\ket{\vacuum}$ is the vacuum state in all spatial mode.
Upon performing an instantaneous measurement, we obtain the following expression:
\begin{equation}
    \begin{aligned}
    \tr\left(\hat{c}^\dag(t)\hat{c}(t) \ket{\psi}_{a} \ket{A_g}_{b} \bra{\psi}_{a} \bra{A_g}_{b}\right) =\\ 
    \tilde{t}^{2} \abs{f(t)}^{2} \sum_{n=0}^{\infty} n \abs{c_{n}}^{2} + \tilde{r}^{2} \abs{A}^{2} \abs{g(t)}^{2} +\\ 
    \tilde{r}\tilde{t} f(t) g^{*}(t) A^{*} \sum_{n=0}^{\infty} c_{n} c^{*}_{n+1} \sqrt{n+1} + \hc,\label{eq:general}
    \end{aligned}
\end{equation}
where the last term is the interference term and couples both the state to be measured and the \gls[hyper=false]{lo}.
For \gls[hyper=false]{kkd}, the \gls[hyper=false]{lo} is assumed to be known, constant, and strong with regard to the signal to be measured, which formally means that $g(t)=1$ up to an amplitude and a phase which can be taken into account in the symbol $A$.
For practicality we write $A=\abs{A}e^{i\theta}$ where $\abs{A}$ is the modulus and $\theta$ is the phase of the local oscillator.
The state to be measured is \gls[hyper=false]{ssb} which means that $f(t)$ is spectrally on the right of the carrier frequency as shown in Fig.~\ref{fig:kk_setup}(b).
Under these conditions, the phase of the last term in Eq.~(\ref{eq:general}) is reconstructed by \gls[hyper=false]{kkd} as it is possible to reconstruct the whole quantity. 
It includes the phase of the state of interest, which is $f(t) \sum_{n=0}^{\infty} c_{n} c^{*}_{n+1} \sqrt{n+1}$, comprising the phase of the spectral mode $\argument{f(t)}$, and the phase of the particle-number statistics $\argument{\sum_{n=0}^{\infty} c_{n} c^{*}_{n+1} \sqrt{n+1}} = \phi$. 
Therefore, the total measured phase here is 
\begin{equation}
    \zeta(t)= \theta - \phi + \argument{f(t)}.
\end{equation}
\par We now examine more specific examples. Consider the phase eigenstate whose wavefunction is
\begin{equation}
  \ket{\psi}_{a} = \left(1 - \abs{z}^{2}\right)^{-1/2} \sum_{n=0}^{\infty} z^{n} \ket{n}_{f},
\end{equation}
with $z = \abs{z} e^{i\phi}$ and $\abs{z} < 1$. 
This state is a normalizable eigenstate of the phase operator~\cite{susskind_quantum_1964} (see Appendix~\ref{app:phase} for a presentation), but is not an eigenstate of the phase operator measured with \gls[hyper=false]{kkd} (see Eq.~(\ref{eq:phase-op})). This example is specifically for justify this difference between the two operators.
For such a state, the phase detected with \gls[hyper=false]{kkd} yields for the interference term:
\begin{equation}
    Af(t) \sum_{n=0}^{\infty} c_{n} c^{*}_{n+1} \sqrt{n+1} = \abs{A}e^{i\theta}f(t) z^{*} \frac{1}{1 - \abs{z}^{2}} \sum_{n=0}^{\infty} \abs{z}^{2n} \sqrt{n+1}.
\end{equation}
Thus, we again obtain the relative phase $\zeta(t)= \theta - \phi + \argument{f(t)}$.
Note that \gls[hyper=false]{kkd} is also effective for reconstructing the phase of superpositions of coherent states, such as the Schrödinger cat state, where the coefficients are given by $c_{n} = \frac{\alpha^{n}}{\sqrt{n!}} (1 + (-1)^{n})$.

\section{Spectral tomography of single photons inspired by Kramers-Kronig detection}\label{sec:single-photon}
In this section, we explore an alternative protocol inspired by \gls[hyper=false]{kkd} that allows for the complete characterization of both the temporal amplitude and phase of single-photon states, that uses spectral engineering. 
To achieve this, the total single-photon state must exist in a coherent superposition of two components: one that satisfies the \gls[hyper=false]{ssb} (spectral superposition) constraint and one that exhibits a quasi-monochromatic spectrum. By combining these two elements, we can effectively capture the full temporal characteristics of the single-photon state. At the detection stage, instead of the PIN photodiode used for intensity measurements, a single-photon detector is employed. This substitution shifts the measurement from intensity to a probability distribution, which provides the necessary mathematical framework to recover the temporal phase of the single-photon state.

\subsection{General idea}\label{sec:single-idea}
We start with a general single-photon state~\cite{blow_continuum_1990}:
\begin{equation}
    \begin{aligned}
        \phket{\psi} &= \hat{a}^\dag\func{\psi}\ket{\vacuum},\\
        &= \int_{-\infty}^\infty \dif t \psi(t)\hat{a}^\dag(t)\ket{\vacuum},
    \end{aligned}
\end{equation}
where $\psi(t)$ denotes a normalized temporal mode. 
Similarly as in the bosonic coherent case, we consider $\psi(t)$ such that it verifies the Hilbert transform relation of Eq.~(\ref{eq:hilbert-transform}):
\begin{equation}
    \psi(t) = A + \chi(t),\label{eq:wavefunction}
\end{equation}
where $A\gg \abs{\chi(t)}$ for all $t$ and $\chi(t)$, the unknown function to measure, is \gls[hyper=false]{ssb}.
Theoretically, this state is not normalizable to one since the monochromatic $A$ contribution has unbounded support. 
In practice, $A$ should be replaced by $A\times \alpha(t)$ where $\alpha(t)$ is a temporal function of very large duration compared to $\chi(t)$ so that, compared to $\chi(t)$, it can be considered to be constant.
It amounts to a function $\alpha(t)$ with localized spectrum.
We stress that the total wavefunction should still be normalized and thus it is quasi-monochromatic because $\abs{A\alpha(t)}\gg \abs{\chi(t)}$.
In this context the measured quantity is $p(t)=\expval{\hat{a}^{\dagger}(t)\hat{a}(t)}$, the probability of measuring a photon at time $t$ on an ideal single-photon detector, and not the intensity measured by a photodiode.
When $A\times\alpha(t)$ is assumed to be constant the time-of-arrival distribution on an ideal single photon detector is
\begin{equation}
    p(t) = \abs{\psi(t)}^2 = \abs{A + \chi(t)}^2.
\end{equation}
The associated noise is not the vacuum quantum fluctuations anymore, but the noise due to the estimation of $p(t)$ with a finite number of measurements.
The idea to reconstruct the phase is thus to plug the estimation of the probability distribution in \gls[hyper=false]{kkd} signal processing and reconstruct the phase of $\psi(t)$ which verifies the Hilbert transform relation:
\begin{equation}
    \begin{aligned}
        \argument{\psi(t)}=-\pval\int_{-\infty}^\infty\dif t'\frac{\ln{\abs{\psi(t')}^2}}{2\pi(t-t')},\label{eq:hilbert_single}
    \end{aligned}
\end{equation}
where the integration limits are supposed to be infinite when the wavefunction verifies Eq.~(\ref{eq:wavefunction}), that is to say when $\alpha(t)$ is almost constant. 
In practice it means that the effective duration taken into account in the integral is smaller than the duration of $\alpha(t)$ and larger than the duration of $\chi(t)$.
Now, the reconstructed phase converges to the phase of $\psi(t)$ as the number of measurements increases. 

\subsection{Proposition of an experimental implementation}\label{sec:single-implem}
\begin{figure*}[ht]
    \centering
    \includegraphics[width=\textwidth]{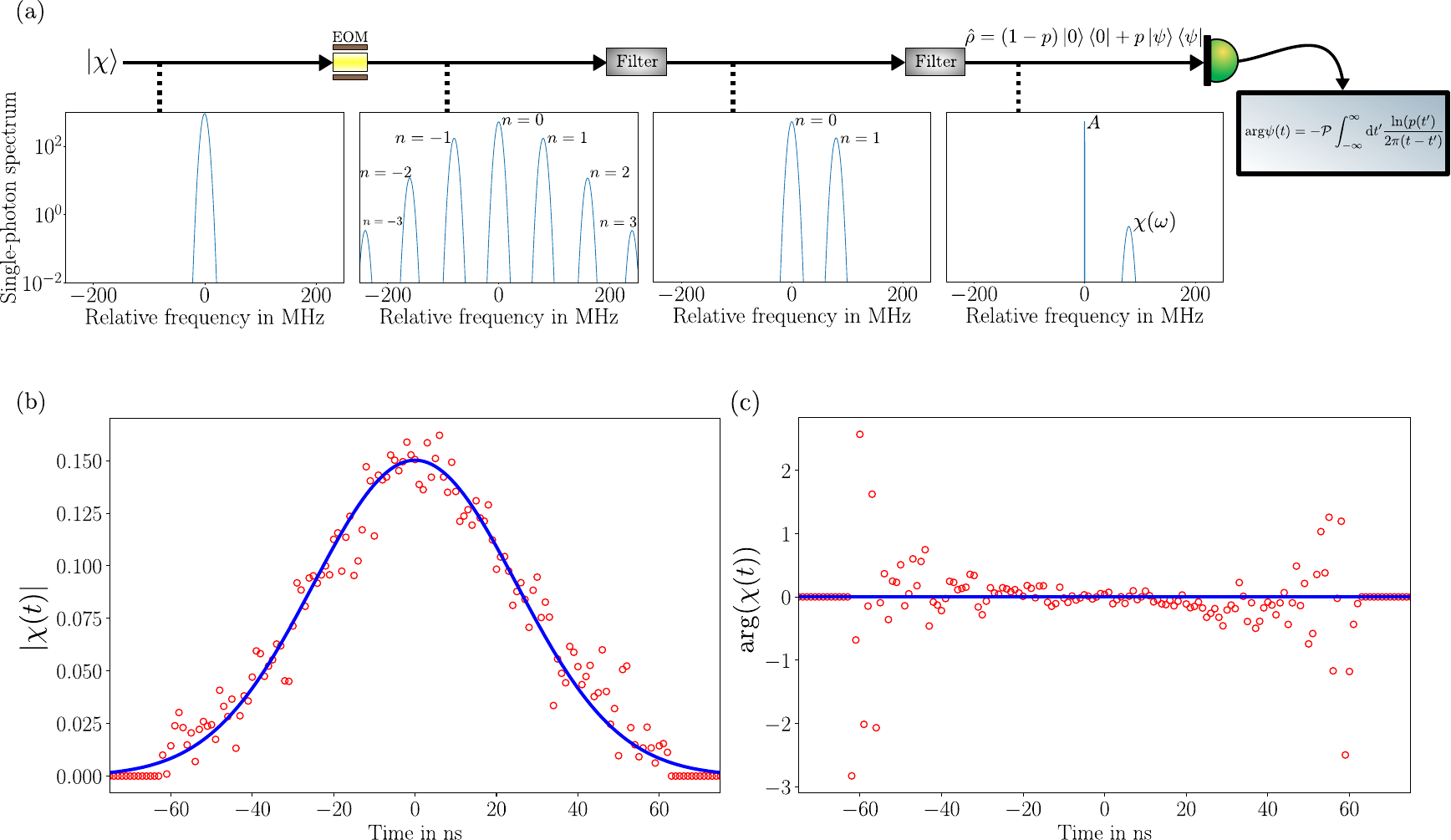}
    \caption{(a) Setup for Kramers-Kronig spectral tomography of the single-photon, (b) simulation results for the reconstruction of the modulus and (c) simulation results for the reconstruction of the phase of a Gaussian wavefunction of bandwidth $40$ MHz using Kramers-Kronig single-photon spectral tomography.
    Kramers-Kronig single-photon spectral tomography (a) consists first in performing signal engineering on the single-photon of interest with the help of an electro optic modulator and frequency filters.
    Below the setup we show the evolution of the spectrum of a Gaussian at each step as an example.
    The result of the signal engineering is a coherent sum of a strong quasi-monochromatic reference and the shifted wavefunction of interest.
    The resulting single-photon time-of-arrival distribution is measured with a single-photon detector.
    The probability undergoes Kramers-Kronig signal processing and the phase is reconstructed.
    The quasi-monochromatic reference is filtered out and the wavefunction of interest is recovered.
    The simulated reconstruction (b,c) accounts for an electro-optic modulator modulated at $80$ MHz with an efficiency of $20\%$.
    The monochromatic reference is created from the copy of order $n=0$ with a lowpass filter of bandwidth $1$ MHz.
    The single-photon detector simulated has a jitter time of $100$ ps and a quantum efficiency of $40\%$.
    A dark count rate of $1$ kHz for a sampling rate of $1$ GHz have been taken into account.
    The plain (blue) curve correspond to the target wavefunction and the round markers (red) correspond to the reconstructed wavefunction.
    The reconstruction reaches a fidelity of $96\%$ for $10^8$ simulated measurements.
    }
    \label{fig:single-setup}
\end{figure*}

In this section, we propose a possible implementation of the \gls[hyper=false]{kkd} single-photon spectral tomography and more particularly of the spectral engineering necessary to turn a single-photon wavefunction $\chi(t)$ in the form of Eq.~(\ref{eq:wavefunction}).
\par The setup proposed is shown in Fig.~\ref{fig:single-setup}(a) and we show the spectrum of the single-photon at each step for a Gaussian wavefunction as an example.
We start with an on-demand or heralded single-photon with temporal wavefunction $\chi(t)$.
The bandwidth $B$ of the single-photon as well as the temporal duration $T$ of the single-photon are supposed to be known such that:
\begin{equation}
    \begin{aligned}
        \chi(t>\frac{T}{2}) \approx \chi(t<-\frac{T}{2}) \approx 0,\\
        \fourier(\chi)(\omega>\frac{B}{2}) \approx \fourier(\chi)(\omega<-\frac{B}{2}) \approx 0.
    \end{aligned}
\end{equation}
The single-photon goes through an \gls[hyper=false]{eom}~\cite{kolchin_2008,capmany_quantum_2011} which modulates the phase periodically with a sine function at frequency $\omega_{\eom}>B$~:
\begin{equation}
    \begin{aligned}
        \chi_{\eom}(t) &= \chi(t)e^{-i\sin(\omega_{\eom}t)},\\
        &= \sum_{n=-\infty}^\infty e^{-in\omega_{\eom} t}\bessel_n(1)\chi(t).
    \end{aligned}
\end{equation}
Here, $\bessel_n$ denotes the Bessel functions of the first kind and we have used the Jacobi-Anger identity~\cite{cuyt_2008} to obtain the second line.
The effect of the phase modulation in frequency is to repeat the spectral wavefunction of interest with the weight of each repetition defined by the Bessel functions of first order.
Since $\omega_{\eom}>B$, the copies do not overlap in frequency domain.
\par Then, we filter out all the repetitions but the ones for $n=0$ which will serve as the quasi-monochromatic reference and $n=1$ which will serve as the \gls[hyper=false]{ssb} signal to reconstruct. 
In the case of an ideal filter of cutoff frequency  $\omega_{\min} = -B/2$ and $\omega_{\max}\in[2\omega_{\eom}-B\slash2,\omega_{\eom}+B\slash2]$ localized between the copy of order $n=1$ and $n=2$ , the input state is projected on the subspace containing only the spectral contribution of the two first copies with probability $p_{\filter}=\abs{J_0(1)}^2+\abs{J_1(1)}^2\approx 0.78$. 
\par The single-photon is filtered again in order to turn the copy of order $n=0$ into a quasi-monochromatic contribution (that plays the role of the LO in this protocol), and leaving the $n=1$ undistorted which is the signal of interest.
The filter for this only leaves a bandwidth $B_{\mathrm{mono}}<B$ of the copy of order $n=0$ and uniformly attenuates the copy $n=1$ by a factor $\gamma$.
As a rule of thumb, assuming that the distribution of the copy $n=0$ of origin is evenly spread in the bandwidth $B$, the state after this second filtering, is a mixed state that can be written as:
\begin{equation}
    \hat{\rho}= (1-p)\ket{0}\bra{0}+p \ket{\psi}\bra{\psi}
\end{equation}
composed of the vacuum $\ket{0}$ and a single-photon $\ket{\psi}$ (whose temporal structure is now in the form of Eq.~(\ref{eq:wavefunction}))  with probability:
\begin{equation}
    p=\abs{J_0(1)}^2B_{\mathrm{mono}}\slash B + \gamma\abs{J_1(1)}^2.\label{eq:ideal-proba}
\end{equation}
This second filter must be tight enough in frequency for the order $n=0$ copy to be considered constant in time compared to the $n=1$ copy.
The attenuation of the repetition $n=1$ must be enough for the quasi-monochromatic reference to be stronger than the repetition $n=1$, to verify the \gls[hyper=false]{mp} constraint.
The single-photon time-of-arrival density probability distribution is measured with a single-photon detector with timing jitter $T_{\mathrm{jit}}$.
As a result the time-of-arrival density probability distribution is sampled at period $t_s$ larger than the jitter time and only approx. $T\slash t_{s}$ histogram points are relevant for the reconstruction.
\par Finally, the reconstructed probability of time-of-arrival undergoes \gls[hyper=false]{kkd} signal processing.
The quasi-monochromatic reference is filtered out with a digital high-pass filter and the initial wavefunction of interest is reconstructed.
\par The first limitations of \gls[hyper=false]{kkd} single-photon spectral tomography regarding the class of single-photon wavefunctions~\cite{kuhn_2002a,mckeever_2004a,farrera_2016} that can be reconstructed concern the bandwidth and the duration of the wavepacket.
The bandwidth $B$ of the single-photon is limited by the modulation speed of the \gls[hyper=false]{eom}, which should be large enough to shift the wavefunction in frequency without causing overlaps.
Bandwidths of \gls[hyper=false]{eom} reach several $\SI{}{GHz}$ in the literature~\cite{wooten_2000,kolchin_2008,koeber_2015,kamada_2022}.
The duration $T$ of the single-photon is limited by the timing jitter of the single-photon detector which should be low enough to have enough samples per pulse for \gls[hyper=false]{kkd} digital signal processing to be efficient.
The timing jitter of current single-photon detectors reaches the picosecond scale in the litterature~\cite{ghioni_2009b,amri_2016,wu_2017,korzh_2020a}.
Finally another constraint is the dark count rate of the single-photon detector which must largely lower-bound the repetition rate of the experiment such that only a low number of parasitic counts are detected compared to the single-photon events.
Current single-photon detectors reach dark count rates at the Hz level~\cite{marsili_2013,amri_2016}.
For each single-shot measurement the probability of success is upper bounded by Eq.~(\ref{eq:ideal-proba}) in the ideal lossless case without dark counts.
\par In Fig.~\ref{fig:single-setup}(b,c) we show a simulation of the reconstruction using our setup idea (see Appendix~\ref{app:kk_single} for a detailed explanation).
The extra losses due to the \gls[hyper=false]{eom} as well as the single-photon detector efficiency are taken into account.
A fidelity of $96\%$ is reached with $10^8$ simulated measurements.
The wavefunction to reconstruct is a Gaussian of characteristic bandwidth $\SI{40}{MHz}$ which corresponds to a duration of $T\approx\SI{100}{ns}$ where $98\%$ of the energy of the gaussian is contained in $[-\frac{T}{2},\frac{T}{2}]$.
The simulated \gls[hyper=false]{eom} is lossy with an efficiency of $20\%$ and a modulation frequency of $\SI{80}{MHz}$.
The filtering process only keeps a centered $\SI{1}{MHz}$ bandwidth from the $n=0$ repetition.
For this value of bandwidth for the quasi-monochromatic reference, an attenuation of $0.04$ has been applied on the $n=1$ repetition.
Finally a lossy single-photon detector of efficiency $40\%$  with a dark count rate of $\SI{1}{kHz}$ and a jitter time of $\SI{100}{ps}$ has been simulated.
For a sampling period $t_s=1$ ns, we have $T\slash t_{s}=100$ bins in the histogram per pulses.
This performances should be reachable for state-of-the-art avalanche photodiodes single photon detectors.

Compared to other protocols in the literature based on shearing and frequency-resolved single-photon detectors ~\cite{legero_characterization_2005,davisExperimentalSinglephotonPulse2018,davisMeasuringSinglePhotonTemporalSpectral2018,thielSinglephotonCharacterizationTwophoton2020,fabreSpectralSinglePhotons2022,kurzynaVariableElectroopticShearing2022a,leiPhaseRetrievalHongOuMandel2024}, the described protocol relies on spectral engineering and time-resolved single-photon detectors.
Given the state-of-the-art regarding single-photon detectors and \gls[hyper=false]{eom}, the described protocol could be used for the spectral tomography of single photons emitted by, \latin{e.g.}, atomic Rubidium clouds~\cite{farrera_2016} or single atoms~\cite{kuhn_2002a,mckeever_2004a} with a bandwidth shorter than that of the single photon detectors.

\section{Conclusion}\label{sec:conclusion}
We have performed the quantization of the \gls[hyper=false]{kkd} detection and calculated the first two-moments of the intensity and phase measured operators with the hypothesis of ideal optics, ideal signal processing and classical \gls[hyper=false]{lo}.
We have shown that \gls[hyper=false]{kkd} is a Gaussian measurement at the first order in the \gls[hyper=false]{lo} meaning that it allows to measure both quadratures of quantum fields at the first order in the \gls[hyper=false]{lo}.
We have also pointed out that \gls[hyper=false]{kkd} yields the same quantum fluctuations as \gls[hyper=false]{dhd}. This detection protocol reconstructs the phase of a quantum state from a direct intensity measurement, however, we stress that the measured \gls[hyper=false]{kkd} phase operator is not the phase operator defined in the quantum optics litterature.
We have illustrated that the phase reconstruction procedure works in particular for bosonic coherent states and mixture of coherent bosonic states, and for monomode pure states. A phase that depends on both the spectral degree of freedom of the mode and on the photon-number statistics can also be reconstructed.
\par Alternatively, we have shown that a protocol based on \gls[hyper=false]{kkd} may be used to perform the spectral tomography of single-photon states.
This estimation of the temporal wavepacket relies on temporally-resolved single-photon detectors and spectral engineering.\\

Furthermore, as a direct detection, \gls[hyper=false]{kkd} shifts the optical complexity of traditional balanced detection into signal processing complexity~\cite{zhong_2018,ma_2024}. 
It would mean that \gls[hyper=false]{kkd} is immune from sources of technical noise related to finite common mode rejection ratio that may limit the performance of quadrature measurement with a balanced coherent receiver for example~\cite{biele_2022}.
In regular balanced detections this issue has to be solved in the optical and electrical domain by fine tuning the experimental setup~\cite{qi_2023} to the balanced photodiodes or the beamsplitters. Finally, direct detections alternatives to \gls[hyper=false]{kkd}, based on iterative algorithms and dispersion for example, have been proposed in the field of classical telecommunications~\cite{randel_100-gbs_2015,li_two-stage_2016,chen_2020,ma_2024}.
In classical regime they act as coherent detectors but their validity could also be investigated in the quantum regime.

\section*{Acknowledgment}
T. Pousset acknowledges funding from IMT, l’Institut Carnot TSN and the Fondation Mines-Télécom. M. Federico ackowledges funding from European Union’s Horizon Europe research and
innovation programme under the project Quantum Secure Network Partnership (QSNP, grant agreement No 101114043). We are also very grateful for fruitful discussions with Elie Awwad.

\appendix

\section{Field quantization and correspondence principle}\label{app:quantization}
In this appendix, we recall the basics of the electromagnetic field quantization that we used to construct the quantum \gls[hyper=false]{kkd} detection from its classical equivalent. 
We will stress the main steps and results but a more detailed approach can be found, e.g., in \cite{glauber_1991,chiao_2008}.
The quantization procedure follows a sequence of steps: 
(a) Starting from Maxwell's equations written in a Hamiltonian form and providing a pair of canonical variables, 
(b) a classical complex Hilbert space $\hilbert$ is built such that it represents the Hamiltonian structure. 
(c) A bosonic Fock space $\fock(\hilbert)=\bigoplus_{n=0}^{\infty} \hilbert^{\otimes_{S} n}$ is constructed, where $\hilbert^{\otimes_S n}$ is the symmetrized $n$-times tensor product of $\hilbert$ with itself. 
Each subspace for a fixed $n$ corresponds then to the subspace of $n$-photon states. 
The passage from one subspace to another is done using bosonic creation-annihilation operators $\hat{a}^\dag\func{\psi}$ and $\hat{a}\func{\psi}$, defined directly for any element $\psi\in\hilbert$, and satisfying the standard bosonic commutation relations $\comm{\hat{a}\func{\psi}}{\hat{a}^\dag\func{\phi}}=\inner{\psi}{\phi}_{\hilbert}$ \cite{blow_continuum_1990,chiao_2008,fabre_modes_2020}. 
(d) The construction of quantum observables is performed using a correspondence principle, initially guided by the harmonic-oscillator-like form of the electromagnetic field Hamiltonian. 
(e) The definition of the generator of the quantum dynamics in Fock space is deduced from the single-photon dynamics, itself determined by the classical Hamiltonian dynamics.
\par In our case, since we use passive optical elements like beamsplitters, one has to consider Maxwell's equations in an inhomogeneous passive dielectric medium characterized by its position-dependent dielectric function.
It has been shown (see, e.g., the pioneering work \cite{glauber_1991}, or more recent contributions \cite{chiao_2008,wubs_2003,federicoSpacetimePropagationPhoton2022}) that a treatment identical to that of the vacuum can be done, the only difference being that the generator of the dynamics will automatically include the medium properties.
Therefore, any real classical field $F(x)$ that can be written in terms of a basis of valid initial conditions for Maxwell's equations, i.e., for any $\{\varphi_\kappa(x)\}\in\hilbert$ as
\begin{align}
F(x)=\sum_\kappa z_\kappa \varphi_\kappa(x) + \cc ,
\end{align}
where $z_\kappa=\braket{\varphi_\kappa}{F}_{\hilbert}$, will be quantized through the following correspondence principle: $z_\kappa\mapsto\hat{a}\func{\varphi_\kappa}$ and $z_\kappa^*\mapsto\hat{a}^\dag\func{\varphi_\kappa}$, which yields the quantum observable
\begin{align}
\hat F(x)=\sum_\kappa \hat{a}\func{\varphi_\kappa} \varphi_\kappa(x) + \hc.
\end{align}
We emphasize that $\kappa$ represents here a possible combination of continuous and discrete indices that correspond to the spectrum degeneracy of the considered system. In particular, one could take, in free space, the basis of circularly polarized plane waves for which  $\kappa\equiv (\vec k,\sigma)$, where $\vec k$ is the wavevector and $\sigma$ denotes the two possible polarizations; for this case, the sum over $\kappa$ in fact hides $\sum_\kappa\equiv\int_{\mathbb{R}^3}d^3k\sum_{\sigma=\pm}$.
\par Similarly as the creation and annihilation operator, to each solution of the Maxwell's equations we associate two hermitian operators called the quadratures~:
\begin{equation}
    \begin{aligned}
        \hat{q}_a\func{\psi} &= \frac{\hat{a}\func{\psi}+\hat{a}^\dag\func{\psi}}{2},\\
        \hat{p}_a\func{\psi} &= \frac{\hat{a}\func{\psi}-\hat{a}^\dag\func{\psi}}{2i}.\label{eq:defquad}
    \end{aligned}
\end{equation}
They follow the canonical position/momentum commutation relations $\comm{\hat{q}_a\func{\psi}}{\hat{p}_a\func{\phi}} = \frac{i}{2}\Re{\braket{\psi}{\phi}_{\hilbert}}$, $\comm{\hat{q}_a\func{\psi}}{\hat{q}_a\func{\phi}} = \comm{\hat{p}_a\func{\psi}}{\hat{p}_a\func{\phi}} = \frac{i}{2}\Im{\braket{\psi}{\phi}_{\hilbert}}$.
In our convention the quadratures are the analog of the real part and the imaginary part of the field in a specified mode which explains the factor $2$ in Eq.~(\ref{eq:defquad}) instead of $\sqrt{2}$.
\par By construction, Maxwell's equations relate vectorial quantities that depends both on time and on the three dimensions of space. 
However, here we did not specify the vectorial dependence since it does not affect the quantization and the coherent detection problems we want to treat can be reduced to one dimensional situations. 
We consider a popagating one dimensional field in the $+x$ direction. 
The only degree of liberty left is the wavevector $k_x\geq 0$, along the propagating direction. 
We apply the change of variable $t=x/c$ and $\omega = c\abs{k_x}$ where the time $t$ is the detection time of the photodiode and $\omega$ is the frequency associated to the spectrum of the signal. 
In the context of optical telecom communications  we consider only wavepackets located around a carrier frequency (typically $1550\unit{nm}$). 
These wavefunctions have low bandwidth compared to the carrier frequency. 
Formally it means that we consider wavefunctions of the form $\psi(t) = C\times e^{-i\omega_c t}f(t)$ where $f(t)$ bandwidth is very small with regard to $\omega_c$ the carrier frequency. 
The constant $C$ being the same for all the family of mode $f$ we consider so that all modes $f$ are normalized to 1. 
In this context we will denote $\hat{a}\func{f}$ and $\hat{a}^\dag\func{f}$ instead of $\hat{a}\func{\psi}$ and $\hat{a}^\dag\func{\psi}$, the frequency shift being implicit.
The inner product used in the commutation relations associated to the considered Hilbert space of modes becomes $\braket{f}{g}_\hilbert = \int_{-\infty}^\infty \dif t f^*(t)g(t)$.
Considering a family of wavefunction $f_\kappa$ we can write the electric field at time $t$:
\begin{equation}
    \hat{E}(t) = Ce^{-i\omega_c t}\sum_\kappa \hat{a}\func{f_\kappa}f_\kappa(t) + \hc.
\end{equation}
Useful operators are the annihilations and creation operators at time $t$: $\hat{a}(t) = \sum_\kappa \hat{a}\func{f_\kappa}f_\kappa(t)$ and $\hat{a}^\dag(t)=\sum_\kappa \hat{a}^\dag\func{f_\kappa}f^*_\kappa(t)$ where $\kappa$ spans over all possible modes.
For any mode $f$, one can express $\hat{a}\func{f}$ and $\hat{a}^\dag\func{f}$ as a function of $\hat{a}(t)$ and $\hat{a}^\dag(t)$:
\begin{equation}
    \begin{aligned}
        \hat{a}\func{f} &= \int_{-\infty}^\infty\dif t f^*(t)\hat{a}(t),\\
        \hat{a}^\dag\func{f} &= \int_{-\infty}^\infty\dif t f(t)\hat{a}^\dag(t).
    \end{aligned}
\end{equation}
In particular, one can define for practical purposes the annihilation/creation operators associated to plane waves at frequency $\omega_c+\omega$:
\begin{equation}
    \begin{aligned}
        \hat{a}(\omega) &= \int_{-\infty}^\infty \dif t e^{i\omega t}\hat{a}(t),\\
        \hat{a}^\dag(\omega) &= \int_{-\infty}^\infty \dif t e^{-i\omega t}\hat{a}^\dag(t),
    \end{aligned}
\end{equation}
with the inverse transform being:
\begin{equation}
    \begin{aligned}
        \hat{a}(t) &= \fourier^{-1}(\hat{a})(t) = \int_{-\infty}^\infty \frac{\dif\omega}{2\pi} e^{-i\omega t}\hat{a}(\omega),\\
        \hat{a}^\dag(t) &= \int_{-\infty}^\infty \frac{\dif \omega}{2\pi} e^{i\omega t}\hat{a}^\dag(\omega),
    \end{aligned}
\end{equation}
where we extended $\omega$ (frequency relative to the carrier) to $-\infty$ for convenience regarding the commutation relations. 
We stress that, $\omega$ is a relative frequency hence the possible negativity.
In practice it means that the considered states are localized around the carrier frequency so that $\omega \rightarrow -\infty$ and $\omega \rightarrow \infty$ plane waves are never populated.
With continuous parametrization $t$ or $\omega$, the operators have to be seen as distribution and their commutation relations yields Dirac distributions~\cite{blow_continuum_1990}: $\comm{\hat{a}(t)}{\hat{a}^\dag(t')}=\delta(t-t')$ and $\comm{\hat{a}(\omega)}{\hat{a}^\dag(\omega')}=\delta(\omega-\omega')$.

\section{Kramers-Kronig signal processing}\label{app:kk-dsp}
In this appendix, we show that the measurement of the intensity leads to the measurement of the phase in the quantum regime from Sec.~\ref{sec:kk}.
We write the input of the photodiode of Fig.~\ref{fig:kk_setup}(a):
\begin{equation}
    \begin{aligned}
        &\tilde{r}\hat{b}(t) + \tilde{t}\hat{a}(t)\\
        &= \tilde{r}A + \tilde{t}\hat{q}(t) + i\tilde{t}\hat{p}(t).
    \end{aligned}
\end{equation}
When calculating the photocurrent at the output of the photodiode we have the quadratic term of the local oscillator, the quadratic term of the signal and cross terms with the signal and the local oscillator. 
These cross terms filter out the imaginary part of the signal since we supposed the local oscillator to be real. 
If the local oscillator had a phase $\theta$ we could write $\tilde{r}\hat{b}(t) + \tilde{t}\hat{a}(t)=e^{i\theta}(\tilde{r}e^{-i\theta}\hat{b}(t) + \tilde{t}e^{-i\theta}\hat{a}(t))$. 
The global phase would be filtered out by the intensity detection and the effective phase space for the signal would be the initial phase space rotated by $\theta$. 
Next we suppose that $\tilde{r}$ and $\tilde{t}$ are real since the phase could be accounted for by a rotation of the phase space of $\hat{a}$. 
The intensity of the photodiode writes:
\begin{equation}
    \hat{I}(t) = \tilde{r}^2A^2 + 2\tilde{t}\tilde{r}A\hat{q}(t) + \tilde{r}^2\hat{a}^\dag(t)\hat{a}(t),
\end{equation}
where the last term is negligible with regard to the other two because of the local oscillator amplitude and the amplitudes of the signal states under study.
To calculate the Hilbert transform of the intensity, one needs to write the logarithm of the intensity as an operator. 
To do so we use a perturbative approach. 
The intensity of the field at the input of the photodiode is equal to zero order to $\tilde{r}^2A^2$.
\begin{equation}
    \begin{aligned}
        \ln{\hat{I}(t)} &=\ln{\tilde{r}^2 A^2\left(1 + \frac{\hat{I}(t)-\tilde{r}^2 A^2}{\tilde{r}^2 A^2}\right)},\\
        &=\ln{\tilde{r}^2 A^2} + \sum_{n=1}^\infty \frac{(-1)^{n+1}}{n}\left(\frac{\hat{I}(t)-\tilde{r}^2 A^2}{\tilde{r}^2 A^2}\right)^n,\\
        &= \ln{\tilde{r}^2 A^2} + \frac{\hat{I}(t)-\tilde{r}^2 A^2}{\tilde{r}^2 A^2} + O\left(\frac{1}{A^2}\right),\label{eq:dl-ln}
    \end{aligned}
\end{equation}
where $O(1\slash A^2)$ denotes an operator whose expectation value for the considered states is of the order of $O(1\slash A^2)$. 
Using the fact that Cauchy's principal value acting on odd functions yields zero, the decomposition in frequency components of the field operator (see Appendix~\ref{app:quantization}) and Dirichlet's integral $\int_{-\infty}^\infty \dif x\sin{x}/x=\pi$, we calculate the phase operator defined by \gls[hyper=false]{kkd}:
\begin{equation}
    \begin{aligned}
        \hat{\varphi}_{\kk}(t) &= -\pval\int_{-\infty}^\infty\dif t' \frac{\hat{I}(t')-\tilde{r}^2A^2}{\tilde{r}^2A^2}\times\frac{1}{2\pi(t-t')} + O\left(\frac{1}{A^2}\right)\\
        &= -\pval\int_{-\infty}^\infty \dif t'\frac{\tilde{t}}{\tilde{r}}\frac{\hat{q}(t-t')}{A\pi t'} + O\left(\frac{1}{A^2}\right),\\
        &= \pval\int_{-\infty}^\infty \dif t'\int_0^\infty \frac{\dif\omega}{2\pi}\frac{\tilde{t}}{\tilde{r}}\left\{\frac{(\hat{p}(\omega)-\hat{p}(-\omega))\cos{\omega t}\sin{\omega t'}}{A\pi t'}\right.\\
        &\left.-\frac{(\hat{q}(\omega)+\hat{q}(-\omega))\sin{\omega t}\sin{\omega t'}}{A\pi t'}\right\}+ O\left(\frac{1}{A^2}\right),\\
        &= \int_0^\infty \frac{\dif\omega}{2\pi}\frac{\tilde{t}}{\tilde{r}}\left\{\frac{(\hat{p}(\omega)-\hat{p}(-\omega))\cos{\omega t}}{A}\right.\\
        &\left.-\frac{(\hat{q}(\omega)+\hat{q}(-\omega))\sin{\omega t}}{A}\right\}+ O\left(\frac{1}{A^2}\right).
    \end{aligned}
\end{equation}
We reconstruct $\hat{a}(t)$ using $\hat{\varphi}_{\kk}(t)$ and the measured intensity~:
\begin{equation}
    \begin{aligned}
        \hat{a}_{\kk}(t) &= \frac{e^{i\hat{\varphi}_{\kk}(t)}\sqrt{\hat{I}(t)}-\tilde{r}A}{\tilde{t}},\\
        &=\frac{\tilde r}{\tilde t}A(1+i\hat{\varphi}_{\kk}(t))\left(1+\frac{\tilde{t}}{\tilde{r}}\frac{\hat{q}(t)}{A}\right) - \frac{\tilde r}{\tilde t}A + O\left(\frac{1}{A}\right)\\
        &=\int_0^\infty\frac{\dif\omega}{2\pi}\left\{(\hat{q}(\omega)+\hat{q}(-\omega))e^{-i\omega t}\right.\\
        &+\left.i(\hat{p}(\omega)-\hat{p}(-\omega)))e^{-i\omega t}\right\} + O\left(\frac{1}{A}\right),\\
        &=\int_0^\infty \frac{\dif\omega}{2\pi} (\hat{a}(\omega)e^{-i\omega t} + \hat{a}^\dag(-\omega)e^{-i\omega t})+ O\left(\frac{1}{A}\right),\label{eq:polar-decomposition}
    \end{aligned}
\end{equation}
where the equality is written to first order in $1/A$ and where the square root has been calculated with a Taylor expansion similar to the logarithm in Eq.~(\ref{eq:dl-ln}): $\sqrt{\hat{I}(t)} = \tilde{r}A\sqrt{1+\frac{\hat{I}(t)-\tilde{r}^2A^2}{\tilde{r}^2A^2}}$.
We stress that the $\hat{a}_{\kk}(t)$ operator is not a bosonic annihilation operator as it does not follow the canonical commutation relations.
\par For a given mode $f(t) = \int_{0}^\infty \frac{\dif\omega}{2\pi} \mathcal{F}(f)(\omega)e^{-i\omega t}$ single-sideband we can estimate $\hat{a}\func{f}$:
\begin{equation}
    \begin{aligned}
        \hat{a}\func{f}_{\kk} &= \int_{-\infty}^\infty\dif t f^*(t)\hat{a}_{\kk}(t),\\
        &=\int_{0}^\infty \frac{\dif\omega}{2\pi} \mathcal{F}(f)(\omega)^*\hat{a}(\omega) \\
        &+ \int_0^\infty \frac{\dif\omega}{2\pi} \mathcal{F}(f)(\omega)^*\hat{a}^\dag(-\omega) + O\left(\frac{1}{A}\right),\\
        &=\hat{a}\func{f} + \hat{a}^\dag\func{f^*} + O\left(\frac{1}{A}\right),
    \end{aligned}
\end{equation}
where $\hat{a}\func{f}_{\kk}$ is the estimator. 
We used Parseval identity for the second equality and the fact that $\mathcal{F}(f^*)(\omega)=\mathcal{F}(f)(-\omega)^*$ for the last equality. 
The quadratures are reconstructed as the hermitian and anti-hermitian parts of $\hat{a}\func{f}_{\kk}$ and we find the result of Eq.~(\ref{eq:quadrature-kk}).

\section{Phase and number operators}\label{app:phase}
In Eq.~(\ref{eq:a_kk}) we used the \gls[hyper=false]{kkd} phase operator as if it was the phase from the polar decomposition of the annihilation operator at the output of the beamsplitter $\hat{c}(t)$, but it is not since it commutes with the intensity operator.
In this appendix, we remind the expression of what is called the phase operator in the quantum optics literature, as well as its eigenstate~\cite{susskind_quantum_1964,vourdas_1993,barnett_1989,levy-leblond_who_1976}, to highlight the differences with the phase measured by \gls[hyper=false]{kkd}.
\par We consider the generic quantum harmonic oscillator, where the ladder operators (also called creation and annihilation operators) can be written as a polar decomposition as proposed by Susskind and Glogover \cite{susskind_quantum_1964,levy-leblond_who_1976,vourdas_1993}
\begin{align}
\hat{a}^{\dagger} = \hat{n}^{1/2} \hat{E}_{+}= \hat{E}_{+} (\hat{n}+1)^{1/2},\\
\hat{a}= (\hat{n}+1)^{1/2} \hat{E}_{-}=\hat{E}_{-} \hat{n}^{1/2},
\end{align}
where
\begin{align}
\hat{E}_{-}=\sum_{n=0}^{\infty} \ket{n}\bra{n+1},\\
\hat{E}_{+}=\sum_{n=0}^{\infty} \ket{n+1}\bra{n}.\\
\end{align}
We have $\hat{E}_{\pm}^{\dagger}=\hat{E}_{\mp}$, thus the operators are not hermitian and are therefore not observables. The expectation value can still be obtained if the full tomography is performed. 
They should be understood as the complex exponentials of the phase operator even though they are not unitary and thus cannot be mathematically written as complex exponentials.
Further, cosine and sine phase operators can be defined: $\hat{C}=1/2 (\hat{E}_{-}+\hat{E}_{+})$, and $\hat{D}=1/(2i) (\hat{E}_{-}-\hat{E}_{+})$
and they are called in such a way since they reproduce the same algebraic structure as the projections of the classical harmonic oscillator state in phase space; however one must be careful as they do not commute since the $\hat{E}_{\mp}$ operators are not exponentials.
Also $\hat{C}$ and $\hat{D}$ are hermitian and therefore are observables. Note that we have the inverse relation:
\begin{equation}
 \hat{E}_{+}=\frac{1}{\sqrt{\hat{n}}} \hat{a}^{\dagger} \ , \  \hat{E}_{-}=\hat{a} \frac{1}{\sqrt{\hat{n}}}.
\end{equation}
The  phase displacement operators can be defined as
\begin{equation}
\hat{D}(\phi)= e^{i\phi \hat{n}}
\end{equation}
and is unitary.
The number displacement operator built from the exponential of the phase operator (increasing and decreasing respectively of $n$ steps):
\begin{equation}
{\cal{\hat{D}}}^{-}(n)=\hat{E}_{-}^{n},\quad {\cal{\hat{D}}}^{+}(n)=\hat{E}_{+}^{n},
\end{equation}
is not unitary but isometric only since the vacuum state does not have a preimage.
This is a consequence of having a half-discrete line for the levels of energy.
\par One can check that:
\begin{equation}\label{labelingcomm}
\comm{\hat{n}}{\hat{E}_{+}}=\hat{E}_{+}, \quad \comm{\hat{E}_{-}}{\hat{n}}=\hat{E}_{-}, \quad \comm{\hat{E}_{+}}{\hat{E}_{-}}= \ket{0}\bra{0}
\end{equation}
which is therefore different than the $e(2)$ algebra due to the "cut", and we have $\hat{E}_{-}\hat{E}_{+}=\mathbb{I}$. 
The Fock state forms a completeness relation:$\sum_{n \in\mathbb{N}} \ket{n}\bra{n}=\mathbb{I}$. 
They are the eigenstates of the number operator $\hat{n}\ket{n}=n\ket{n}$, that can be written as $\hat{n}=\sum_{n=0}^{\infty} n\ket{n}\bra{n}$.
\par A correct (normalizable) eigenstate of the exponential of the phase operator $\hat{E}_{-}$, $\hat{E}_{-}\ket{z}=z\ket{z}$ is defined as follows \cite{vourdas_1993,levy-leblond_who_1976}
\begin{equation}\label{eq:phasestate}
\ket{z}=(1-\abs{z}^{2})^{-1/2} \sum_{n=0}^{\infty} z^{n} \ket{n}, 
\end{equation}
with $\abs{z}<1$. 
Note that the mode structure is implicit in the notation $\ket{n}$ that should be read as $\ket{n}_{\psi}$. 
Finally, the phase state can appear as a coherent state in a truncated Fock space~\cite{garciadeleon_2007}. 
Another approach is the Pegg and Barnett formalism~\cite{barnett_1989} where the Fock space is truncated up to a maximum number of excitation, and the limit towards infinity is taken at the end of every expectation value calculations.
If $\abs{z}=1$, the state is not normalizable : it would be convenient to have such a form as a Fourier series, but this is not the case. 
The truncated phase eigenstate can be written as $\ket{\phi}=(\frac{1}{s+1})^{1/2} \sum_{n=0}^{s} e^{i\phi n}\ket{n}_{\psi}$, from which we do not have constraints on the radius anymore for having a normalized state.
Notice that if one takes the limit $s\rightarrow\infty$, the state is not normalizable anymore and corresponds to the eigenstate mentioned in~\cite{susskind_quantum_1964,levy-leblond_who_1976}.
\par The phase operator introduced in this paper necessary commutes with the number operator as $\hat{\varphi}_{\kk}(t)$ is a function of $\hat{n}(t)$:
\begin{equation}
    \comm{\hat{n}(t')}{\hat{\varphi}_{\kk}(t)}=0,
\end{equation}
and even after integrating over time, $\hat{\varphi}_{\kk}(t) $ is not the logarithm of $\hat{E}_{\pm}$ because it is not well defined as it is singular. 
Therefore, the eigenstate of the exponentional of the phase operator $\hat{E}$ is not the one of $\hat{\varphi}_{\kk}(t)$.
\par Another naive attempt to define a phase operator would be by adding a filtering function $v(t)$ as it is done for the annihilation operator:
\begin{equation}\label{eq:mode}
 \begin{aligned}
     \hat{\varphi}_{\kk}[v]&=\int\dif t v(t) \hat{\varphi}_{\kk}(t) =\iint\dif t\dif t'\frac{1}{\pi(t-t')} v(t)\ln(\hat{I}(t')) \\
    &= \int \dif t' f(t')\ln(\hat{I}(t'))= \ln(\hat{I})[f]
 \end{aligned}
\end{equation}
where $f(t')=\int\dif t \frac{v(t)}{2\pi(t-t')} $. 
However, it is certain that it can not respect the relation $\comm{\hat{n}}{e^{i\hat{\varphi}_{\kk}}}=e^{i\hat{\varphi}_{\kk}}$ as in Eq.~(\ref{labelingcomm}).
Indeed, in Eq.~(\ref{eq:mode}), the phase and the intensity are projected into different modes.

\section{Average value of the phase operator with respect to mixed state}\label{app:mixedstate}
In this appendix, we present the main calculations for obtaining the expression of the average value of the phase measured with \gls[hyper=false]{kkd}, from the moments of the direct intensity operator.
The expectation value of the instantaneous intensity operator is:
\begin{multline}
\Tr(\hat{\rho}\hat{a}^\dag(t)\hat{a}(t))=
\\\iint\Dif\func{\alpha}\Dif\func{\beta}\pfunc\func{\alpha}\abs{\braket{\func{\beta}}{\func{\alpha}}}^{2} \alpha^{*}(t) \beta(t)
\end{multline}
where we have used the completeness relation of the bosonic coherent state $\int \Dif \func{\alpha}\cohtket{\alpha}\cohtbra{\alpha}= \mathbb{I}$. The scalar product is $\abs{\braket{\func{\beta}}{\func{\alpha}}}^{2}=\exp(-\int\dif t\abs{\beta(t)-\alpha(t)}^{2})$ (see \cite{roux_2020,robertCoherentStatesApplications2021}). 
The integration over $\beta$ leads to:
\begin{equation}
\int\Dif\func{\beta}\exp(-\int\dif t'\abs{\beta(t')-\alpha(t')}^{2})\beta(t)=\alpha(t)
\end{equation}
where to perform the functional integral, we can perform the derivative with respect to $\alpha^{*}(t)$), or by using a change of variable. 
Finally, we obtain:
\begin{equation}
\Tr(\hat{\rho}\hat{a}^\dag(t)\hat{a}(t))=\int\Dif\func{\alpha}\pfunc\func{\alpha} \abs{\alpha(t)}^{2}
\end{equation}
The average value of the phase obtained with \gls[hyper=false]{kkd} puts into play the expression of the logarithm of the intensity, logarithm that is expanded into a power series, with respect to a strong local oscillator.
Thus every moment of the intensity operator appears.
We have for instance:
\begin{equation}
\Tr(\hat{\rho}(\hat{a}^\dag(t)\hat{a}(t))^{2})= \int\Dif\func{\alpha}\pfunc\func{\alpha}( \abs{\alpha(t)}^{4}+\abs{\alpha(t)}^{2}).
\end{equation}
Using the result of~\cite{mansour_2015} for the n-th power we compute the logartihm of the intensity operator:
\begin{multline}
    \Tr{\hat{\rho}\ln{\hat{a}^\dag(t)\hat{a}(t)}}=\\
    \ln{A^2}+\sum_{n=0}^\infty (-1)^{n-1} \frac{\Tr{\hat{\rho}(\hat{a}^\dag(t)\hat{a}(t)-A^2)^{n}}}{nA^{2n}},\\
    =\ln{A^2}+\sum_{n=0}^\infty\sum_{m=0}^n\frac{(-1)^n}{nA^{2n}}\binom{m}{n}A^{n-m}\Tr{\hat{\rho}(\hat{a}^\dag(t)\hat{a}(t))^n},\\
    =\ln{A^2}+\sum_{n=1}^\infty\sum_{m=1}^n\frac{(-1)^n}{nA^{2n}}\binom{m}{n}A^{n-m}\sum_{j=1}^m\stirling{j}{m}\Tr{\hat{\rho}\hat{a}^\dag(t)^n\hat{a}(t)^n},
\end{multline}
where $\stirling{j}{m}$ denotes the Stirling numbers of the second kind~\cite{mansour_2015}.

\section{KK single photon tomography}\label{app:kk_single}

In this section, we describe the model used in the numerical simulation for Kramers-Kronig spectral tomography of single photons (see Fig.~\ref{fig:single-setup}). The simulations account for losses from both the electro-optic modulator (EOM) and the single-photon detector, as well as noise contributions from dark counts, timing jitter, and statistical fluctuations due to the finite number of measurements.\\

We consider the spectral tomography of single-photon states generated by heralded atomic spontaneous parametric down conversion sources or through cavity-stimulated Raman adiabatic passage, as these sources have a larger temporal wavepacket compared to the typical single-photon detector temporal resolution. This requirement cannot be met by integrated or bulk sources, which typically exhibit temporal widths on the order of picoseconds or femtoseconds. As for the detectors, we will consider reconstruction with both avalanche photodiodes (APD) and superconducting nanowire single-photon detectors (SNSPD), to demonstrate the possibility of the spectral reconstruction with various detectors quantum efficiencies.

\subsection{Time sampling and time's truncation}
In our numerical simulations, we model the continuous-time wavefunction of the single-photon state at each stage of the setup, \(\chi(t)\), as a discrete finite series, \(\tilde{\chi}(n)\), where \(n\) spans from \(-N/2\) to \(N/2\), with \(N+1\) representing the total number of time samples.  

The wavefunction in our simulations is given by:  
\begin{equation}
\tilde{\chi}(n) = N_{\chi} e^{-\frac{n^2 t_s^2 B^2}{2}},
\end{equation}
where \(N_{\chi}\) is a normalization factor, \(t_s\) is the sampling period (set to 1 ns), and \(B\) is the bandwidth of the single-photon (taken as 40 MHz).  

This wavefunction extends over approximately 25 ns, a duration that should be feasible for atomic sources \cite{kuhn_2002a, mckeever_2004a, farrera_2016}. Given the chosen bandwidth and sampling period, around \(n \approx 200\) samples are sufficient to capture most ($99.9\%$) of the Gaussian’s energy. However, to achieve better frequency resolution, we increase the total number of samples to \(N \approx 5000\) by padding the numerical wavefunction with zeros before and after the main signal.

\subsection{Model of the electro-optic modulator}

The model for the electro-optic modulator (EOM) accounts for both linear phase modulation and losses. The EOM applies a time-dependent phase shift to the input signal $\exp(i\varphi_{\mathrm{EOM}}(t))$, where $\varphi_{\mathrm{EOM}}(t) = \pi V(t)/V_\pi$. Here, \(V(t)\) is the applied voltage, and \(V_\pi\) is the half-wave voltage of the EOM.

For our simulation, we assume an ideal EOM with infinite linear bandwidth and amplitude range. The applied voltage is defined as:  
\begin{equation}
V(t) = -\frac{V_{\pi}}{\pi} \sin(\omega_{\mathrm{EOM}} t),
\end{equation}
corresponding to a modulation index of \(-1\), and we will set for the numerical simulation $\omega_{\mathrm{EOM}}=80 $MHz.  

We set for the EOM a 7 dB loss, which we model as an attenuation factor \(\gamma_{\mathrm{EOM}}^{-1}\) on the detection probability, translating to an amplitude attenuation factor of \(\gamma_{\mathrm{EOM}}^{-1/2}\). These characteristics should be achievable with current electro-optic modulators \cite{wooten_2000, koeber_2015, kamada_2022}. Therefore, the EOM output is a discrete signal $\tilde{\chi}_{\mathrm{EOM}}(n)$ that can be written as:
\begin{equation}
    \begin{gathered}
        \tilde\chi_{\mathrm{EOM}}(n) = \gamma_{\mathrm{EOM}}^{-1/2}\tilde\chi(n)e^{i\pi\frac{V(nt_s)}{V_{\pi}}}\\
        =\gamma_{\mathrm{EOM}}^{-1/2}\tilde\chi(n)e^{-i\sin(\omega_{\mathrm{EOM}}nt_s)}\\
        =\gamma_{\mathrm{EOM}}^{-1/2} \tilde\chi(n) \sum_{k=-\infty}^\infty J_k(-1)e^{-ik \omega_{\mathrm{EOM}}nt_s},
    \end{gathered}
\end{equation}
where we have used $e^{iz\text{sin}(\phi)}=\sum_{n} J_{n}(z) e^{in\phi}$, and $J_k$ denotes the $k$-th Bessel function of the first kind~\cite{cuyt_2008}.
The output signal is composed of non-overlapping aliases of the original wavefunction shifted by multiples of $80$ MHz in frequency domain and weighed by the Bessel functions of first order.

\subsection{Filtering operation}

The filtering operation first removes unnecessary copies of the wavefunction generated after the EOM, retaining only the \( k = 0 \) and \( k = 1 \) components.  The second function of the filtering process is to transform the \( k = 0 \) contribution into a quasi-monochromatic signal while attenuating the \( k = 1 \) copy, making it significantly weaker than the quasi-monochromatic component.  Numerically, the first step is performed in the Fourier domain by zeroing out all frequency components outside the range \([-B/2, B/2 + \omega_{\mathrm{EOM}}]\). The resulting output in the Fourier domain is given by
\begin{equation}
    \mathcal{F}\{\tilde{\chi}_{\mathrm{filter1}}\}(\omega) = \left\{\begin{array}{lll}
    0 &\omega < -\frac{B}{2},\\
\mathcal{F}\{\tilde\chi_{\mathrm{EOM}}\}(\omega) &
 -\frac{B}{2}<\omega<\omega_{\mathrm{EOM}}+\frac{B}{2},\\
    0 &\omega>\omega_{\mathrm{EOM}}+\frac{B}{2}.
    \end{array}\right.
\end{equation}
The second stage of the filtering process refines the remaining \( k = 0 \) component, reducing its bandwidth to \( B_{\mathrm{mono}} = 10^6 \). Simultaneously, the \( k = 1 \) copy is attenuated by a factor of \( \gamma_{\mathrm{filter}}^{-1/2} = 1/25 \).  The resulting output in the Fourier domain is given by:
\begin{align}
    \mathcal{F}\{\tilde{\chi}_{\mathrm{filter2}}\}(\omega) = \left\{\begin{array}{lll}
        0&\omega<-\frac{B_{\mathrm{mono}}}{2},\\
        \mathcal{F}\{\tilde\chi_{\mathrm{filter1}}\}(\omega)&-\frac{B_{\mathrm{mono}}}{2}<\omega<\frac{B_{\mathrm{mono}}}{2},\\
        \gamma_{\mathrm{filter}}^{-1/2}\mathcal{F}\{\tilde\chi_{\mathrm{filter1}}\}(\omega)& \frac{B_{\mathrm{mono}}}{2}<\omega<\omega_{\mathrm{EOM}}+\frac{B}{2}.
    \end{array}\right.
\end{align}

\subsection{Detector jitter time}

The single photon detector model in the simulation incorporates timing jitter~\cite{legero_characterization_2005}. This jitter acts as a temporal filter, \( f_{\mathrm{jitter}}(t) \), with duration \( t_{\mathrm{jitter}} \), applied to the time-of-arrival probability \( p(t) \) that would be observed without jitter. In the continuous time domain, the jitter-affected probability is given by:
\begin{equation}
    p_{\mathrm{jitter}}(t) = \int_{-\infty}^{\infty} \mathrm{d}\tau \, f_{\mathrm{jitter}}(t-\tau) \, p(\tau).
\end{equation}
In our simulation, the jitter filter is modelled as a Gaussian with a duration \( t_{\mathrm{jitter}} = 0.1 \) ns and is implemented numerically using the \texttt{gaussian\_filter} function from the SciPy library. In discrete time, the jitter-affected time-of-arrival probability is expressed as:
\begin{equation}
    \tilde{p}_{\mathrm{jitter}}(n) = \left(\tilde{f}_{\mathrm{jitter}} * \tilde{p}\right)(n),
\end{equation}
where \( \tilde{f}_{\mathrm{jitter}} \) is a discrete Gaussian filter truncated at four times its standard deviation, \( \sigma = t_{\mathrm{jitter}}/t_s \), and \((f * g)(n)\) denotes the discrete convolution between \( f \) and \( g \).

\subsection{Time-of-arrival sample generation and losses}
For each emission the density of probability of the time-of-arrival is measured with the single-photon detector. These times are sampled to the sampling period $t_s$ and are put in a histogram, which approximates the time-of-arrival probability $p_{\mathrm{jitter}}(t)$.
\par Numerically, we genrate samples of indices $n$ following the probability distribution $\tilde{p}_{\mathrm{jitter}}(n)$. 
This is done by generating a random uniform number $x$ between zero and one and taking the index $j$ such that $\sum_{j=-N/2}^{n-1}p_{\mathrm{jitter}}(j)<x$ and $\sum_{j=-N/2}^{n}p_{\mathrm{jitter}}(j)>x$.
If there are no such $n_{\mathrm{det}}$ then we assume that no photon has been measured.
The index $n_{\mathrm{det}}$ corresponds to the time interval $[n_{\mathrm{det}}t_s,(n_{\mathrm{det}}+1)t_s]$ at which a single-photon would have been measured.
To account for the losses $\eta$ we multiply the time-of-arrival probability by $\eta<1$, increasing thus the probability that no index $n_{\mathrm{det}}$ satisfy the above mentioned conditions, and thus that no photon is measured.
The probability of finding no $n_{\mathrm{det}}$ matching the conditions correspond in the end to the total loss of the experiment.
Finally all the samples are gathered and a histogram approximating $\tilde{p}_{\mathrm{jitter}}(n)$ is obtained.

\subsection{Dark count model}

Additionally to the time sampling of the single-photon density probability, dark counts have been taken into account using a Poissonian statistics model, as a first approximation~\cite{vinogradov_2009,tzou_2015,menkart_2022}.
For a more precise model of the dark counts, single-photon detectors dead time must be taken into account as shown in Ref.~\cite{menkart_2022}. 
For the model we use, the probability of having no dark count in a time slot $t_s$ is $e^{-rt_s}$, where $r$ is the dark counts rate.
Therefore, the probability of dark count occurring during each time slot $t_{s}$ is:
\begin{equation}
    p_{\mathrm{dc}}=1-e^{-rt_s}\approx rt_s
\end{equation}
where the approximation is based on the fact that the sampling rate $1$ GHz is much larger than the typical dark count rate, i.e., $1$ kHz for APDs.
At each iteration of the simulation, each slot undergoes a Bernoulli test to generate the dark counts.
For each repetition of the experiment, we generate an index $n_{\mathrm{det}}$ corresponding to the single-photon detection event, as well as a list of indices $(n_{\mathrm{dc},1},n_{\mathrm{dc},2},\hdots)$ corresponding to the dark counts.
We add a single-count to the histogram at each detected index even in the case where $n_{\mathrm{det}}$ and one of the $(n_{\mathrm{dc},1},n_{\mathrm{dc},2},\hdots)$ are the same.
Consequently, events with multiple clicks are not discarded, even if one arises from a dark count.

\subsection{Simulation results}

The total histogram with both the photon detection events and the dark count events is normalized and undergoes KKD digital signal processing.
With the phase information, the wavefunction just before the detection $\tilde\chi_{\mathrm{filter2}}(k)$ is reconstructed.
The reconstruction is filtered with a low-pass filter to remove the contribution of the strong monochromatic reference.
Then the reconstructed wavefunction $\tilde\chi_{\mathrm{rec}}(k)$ is frequency-shifted and renormalized to match the original single-photon wavefunction.
The fidelity between the reconstructed wavefunction and the original is given by:
\begin{equation}
    \mathrm{F}=\abs{\int_{-\infty}^{\infty}\mathrm{d}t\chi(t)\chi_{\mathrm{rec}}(t)}^2\approx\abs{\sum_{n=-N/2}^{N/2}\tilde\chi(n)\tilde\chi_{\mathrm{rec}}^*(n)}^2.
\end{equation}
The Fig.~\ref{fig:single-setup}(b),(c) is an example of reconstructed wavefunction modulus and phase in the temporal domain. 
In this simulation, $10^8$ have been taken into account.
The detector has a jitter time of $0.1$ ns, a quantum efficiency of $40\%$, as well as a dark count rate of $1$ kHz, which are typical values for APDs.
The fidelity of the reconstruction in this case is $96\%$.\\

\begin{figure}
    \centering
    \includegraphics[width=\columnwidth]{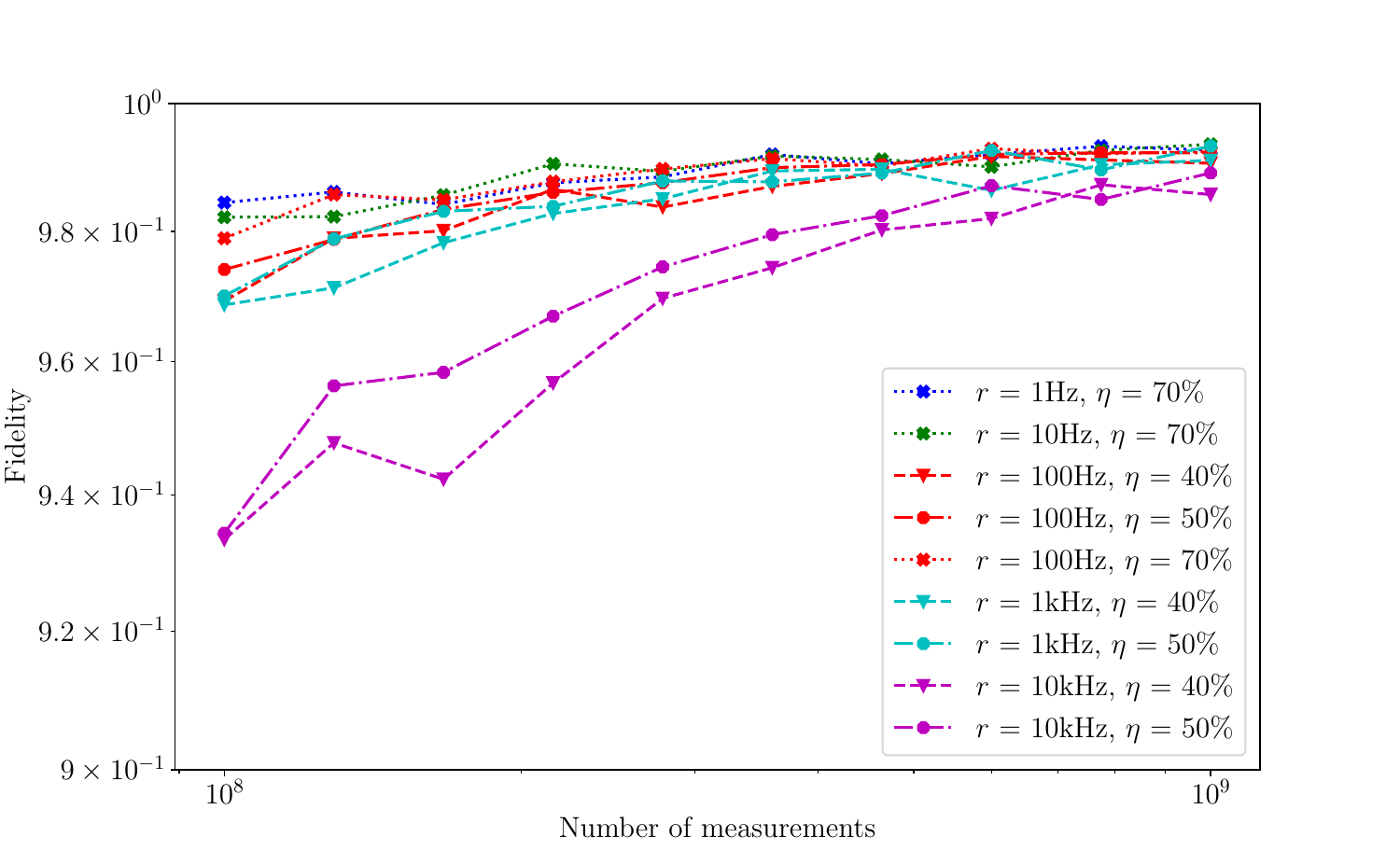}
    \caption{
    Fidelity versus number of measurements taking into account all the clicks of the photodetector in the histogram reconstruction.
    $\eta$ denotes the efficiency of the detector used in the simulation and $r$ denotes the dark counts rate.
    In the simulations, the sampling rate is set to $1$GHz and the jitter of the photodetector is modelled by a Gaussian of duration $0.1$ns.
    For lower efficiency and higher dark count rates, more repetition of the experiments is needed to reach a target fidelity.
    }
    \label{fig:no-disc}
\end{figure}

In Fig.~\ref{fig:no-disc}, we plot the fidelity of the reconstruction as a function of the number of simulated experiments, \( N_{\mathrm{exp}} \), in the case where all detection events are considered, even when an iteration records more than two clicks.  We analyze different photodetector efficiencies \( \eta \) and dark count rates \( r \), selecting values representative of commonly used detectors: APDs with \( \eta \approx 40\% \) and SNSPDs with \( \eta \approx 70\% \).  For APDs, where \( \eta = 40\% \) and \( r \) is on the order of \( 1 \) kHz, approximately \( 3 \times 10^8 \) measurements are sufficient to reach a fidelity of 98\%. In contrast, for SNSPDs with \( \eta = 70\% \) and \( r \) around \( 1 \) Hz, \( 10^8 \) measurements appear to be enough to achieve the same fidelity.  However, at a high number of measurements, the fidelity appears to plateau, suggesting a limiting factor beyond statistical noise. This limitation may stem from the nonzero dark count rate, which introduces additional noise in the measurements that does not diminish with increasing experiment repetitions.

\bibliography{bibliography}
\end{document}